\documentclass[useAMS,usenatbib]{mn2e}
\usepackage{graphicx}
\usepackage{longtable}
\usepackage{lscape}

\def\lx{$L_{{\rm X}}$}
\def\lk{$L_{{\rm K_{{\rm S}}}}$}
\def\lb{$L_{{\rm B}}$}
\def\lh{$L_{{\rm H}}$}
\def\lbsun{$L_{{\rm B}\odot}$}

\def\lksun{$L_{{\rm K_{{\rm S}}}\odot}$}
\def\lhsun{$L_{{\rm H}\odot}$}

\title[Correlations of near-infrared, optical and X-ray luminosity for early-type galaxies]{Correlations of near-infrared, optical and X-ray luminosity for early-type galaxies}
\author[S. C. Ellis and Ewan O'Sullivan]{S. C. Ellis$^{1}$\thanks{E-mail: sce@aao.gov.au} and Ewan O'Sullivan$^{2}$\\
$^{1}$ Anglo-Australian Observatory, P.O. Box 296, Epping, NSW 2121, Australia\\
$^{2}$ Harvard Smithsonian Center for Astrophysics, 60 Garden Street, Cambridge, MA 02138, USA}

\begin{document}

\date{Accepted........Received..........; in original form..........}


\maketitle

\label{firstpage}

\begin{abstract}
The relation between X-ray luminosity and near-infrared luminosity for early-type galaxies has been examined.  Near-infrared (NIR) luminosities
should provide a superior measure of stellar mass compared to optical luminosities used in previous studies, especially if there is significant star-formation or dust
present in the galaxies.  However, we show that the X-ray-NIR relations are remarkably
consistent with the X-ray-optical relations.  This indicates that the large scatter of the relations is dominated by scatter in the X-ray
properties of early-type galaxies, and is consistent with early-types consisting of old, quiescent stellar populations.  

We have investigated scatter in terms of environment, surface brightness profile, Mg$_{2}$, H$\beta$, H$\gamma$ line strength indices, spectroscopic age, and nuclear H$\alpha$ emission.  We found that galaxies with high Mg$_{2}$ index, low H$\beta$ and H$\gamma$ indices or a `core' profile have a large scatter in \lx, whereas galaxies with low Mg$_{2}$, high H$\beta$ and H$\gamma$ indices or `power-law' profiles, generally have \lx$<10^{41}$ erg s$^{-1}$.  There is no clear trend in the scatter 
with environment or nuclear H$\alpha$ emission.

\end{abstract}


\section{Introduction}

Early-type galaxies are known to emit X-rays via discrete populations of low-mass X-ray binaries and via a hot
thermal plasma (\citealt{fab84}; \citealt{for85}; \citealt{fab89}).  The correlation of X-ray luminosity (\lx) and \emph{B}-band optical luminosity (\lb)
has been studied to examine the variation of gas properties as a function of galaxy mass (see e.g.\ 
\citealt{osul01}; \citealt{mat03} for a review).  It is found that there is a transition in the X-ray emission, and hence the hot gas content, of
galaxies at \lb$\approx3 \times 10^{10}$\lbsun.  Galaxies less
luminous than this threshold have X-ray emission dominated by discrete sources, and display a relation
\lx$\propto$\lb.  Above this threshold, there is an excess of X-ray emission attributed to hot gas emitting via
thermal Bremsstrahlung, and hence a steeper relation of \lx$\propto$\lb$^{\sim 2}$ (\citealt{osul01}).

The scatter of the \lx:\lb\ relation is very large, varying by $\sim2$ orders of magnitude for galaxies of similar
\lb\ (\citealt{mat03}).  Several physical phenomena have been examined to account for this scatter.  \citet{mat03}
list: environmental and intrinsic galaxy properties (\citealt{whit91};  \citealt{esk95a}; \citealt{esk95b}; \citealt{esk95c};
 \citealt{hen99}; \citealt{bro00}), differences between power-law ellipticals and core
ellipticals due to their different location within groups and the additional contribution of intragroup gas to centrally
located ellipticals (\citealt{pel99}; \citealt{hel01}; \citealt{mat01}), the influence of rotation and flattening
(\citealt{nul84}; \citealt{kle95}; \citealt{brig96}; \citealt{pel97}; \citealt{der98}), the influence of type Ia
supernov\ae\ and cooling flows (\citealt{cio91}), the extent of the diffuse gas (\citealt{mat98}) and ram-pressure
stripping (\citealt{ton01}).  Recently \citet{pip05} have shown that a late secondary accretion of gas can account for the \lx:\lb\ relation of massive galaxies, and the large observed scatter.

These phenomena play a role in contributing to the large scatter in the \lx:\lb\ relation, but there is often
conflicting evidence and/or too little scatter explained.  

All the phenomena above relate to an increase in
scatter of the X-ray luminosity.  The assumption that the scatter is due to variance in \lx\ rather than \lb\ is
reasonable, since the stellar populations of early-type galaxies are generally thought to be old, passively
evolving systems, especially for galaxies in clusters (e.g.\ \citealt{sta98}; \citealt{dep99}; \citealt{bel04}; \citealt{hol04}; 
\citealt{ell04}).  However, some authors find that there is some evolution of early-type 
galaxies not attributable to purely passive evolution (e.g.\ \citealt{but84}; \citealt{dre97}; \citealt{dro03}; \citealt{vdv03}; \citealt{red04}) in agreement with models of structure formation which predict a merger origin for elliptical galaxies
(e.g.\ \citealt{bau96}).

If elliptical galaxies are not all old and quiescent then there may be significant inconsistencies between the
optical \emph{B}-band luminosity and the mass of a galaxy.  The presence of young stars can significantly enhance the blue
luminosity of a galaxy, and hence the \emph{B}-band would yield an inaccurate estimate of galaxy mass in the presence of
young stellar populations.  Thus, the assumption that the \emph{B}-band luminosity contributes an insignificant scatter 
to the \lx:\lb\ relation needs to be tested.

\begin{figure}
\centering \includegraphics[angle=0,scale=0.4]{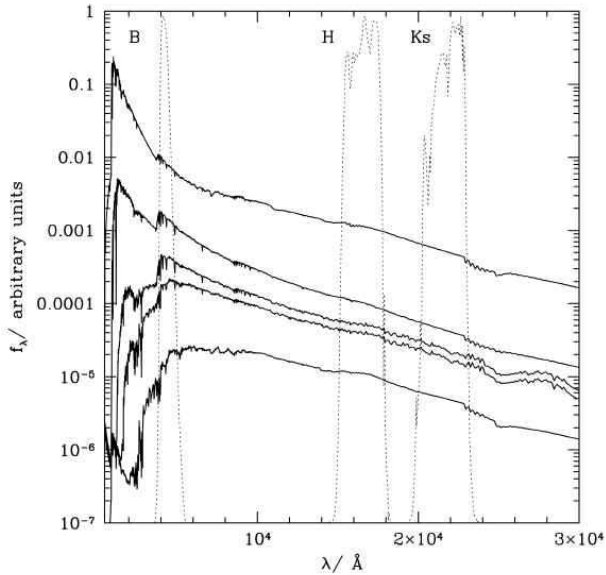}
\caption{Model spectral energy distributions from the libraries of \protect\citealt{bru03}.  The spectra are 10Myr, 100Myr, 500Myr, 1Gyr and 9Gyr old, with the younger spectra being brighter.   The units of flux are arbitrary.  Also illustrated are the locations of the transmission profiles of the \emph{B}, \emph{H} and \emph{Ks} bands. }
\label{fig:examplespecs}
\end{figure}

Near-infrared (NIR) luminosities are much more closely correlated with dynamical and total galaxy mass than optical
luminosities (\citealt{gav96}), since NIR emission is dominated by longer lived stellar populations, and has
little contribution from young, bright stars.  Figure~\ref{fig:examplespecs} shows the model spectral
energy distribution of a simple stellar population at ages, 10Myr, 100Myr, 500Myr, 1Gyr and 9Gyr, from the libraries of \citet{bru03}.  Also shown are the
transmission profiles of the \emph{B}, \emph{H} and \emph{Ks} bands.  The difference between the two stellar populations due to the
massive, short lived stars will clearly have a much stronger effect on the \emph{B}-band luminosity than the \emph{H} or \emph{Ks} 
luminosities, e.g.\ between the ages of 100Myr and 9Gyr, the \emph{B} band luminosity of a simple stellar population decreases by $\approx 14$ times more than the decrease in the  \emph{K} band luminosity, or $\approx 5$ times between 1Gyr and 9Gyr.

The NIR is also much less affected by the presence of dust.  Galactic extinction in the \emph{K} band is typically a factor of $\sim0.08$ times the \emph{B}-band extinction (\citealt{schl98}).  Thus if some elliptical galaxies contain significant amounts of dust the NIR will be a better indicator of the underlying stellar mass.  About 80\% of early-type galaxies contain dust, though this is mainly confined to the central kpc (\citealt{mat03}).  Theoretical work suggests that more widely distributed dust will be rapidly sputtered (e.g.\ \citealt{tsa95}), therefore the presence of dust is unlikely to affect global NIR$-$optical colours.  However, more extended filaments of dust have been observed in some galaxies (\citealt{tra01}).  Any effect due to dust, though expected to be small, will be much reduced in the NIR.

The improved measure of galaxy mass attainable through NIR photometry provides an opportunity to recast the
\lx:\lb\ relation as an \lx:\lk\ relation (or \lx:\lh).  Doing so will test the importance of variance in \lb\ in contributing to the scatter of the \lx:\lb\ relation.
In turn the comparison of the \lx:\lb\ and \lx:\lk\
relations may yield information on the stellar populations of elliptical galaxies, in particular the presence of
young stars.  Accordingly we have reanalysed the X-ray catalogue of early-type galaxies of \citet{osul01}, supplemented with \emph{H} and \emph{Ks} band luminosities from the Two Micron All
Sky Survey (2MASS).

The paper is organised as follows.  Section~\ref{sec:samp} introduces the X-ray catalogue of \citet{osul01}, and
section~\ref{sec:nir} describes the NIR data.  Section~\ref{sec:method} describes the statistical analyses used to fit
the relations.  Section~\ref{sec:results} describes the correlations, and the new
measurements of the scatter as a function of environment, surface brightness profile, Mg$_{2}$, H$\beta$ and H$\gamma$ emission line strength and nuclear H$\alpha$ emission.  Finally section~\ref{sec:discussion} discusses the results in terms of the origin of
the scatter and the stellar populations of early-type galaxies.

\section{Sample}
\label{sec:samp}

\citet{osul01} present a catalogue of X-ray luminosities for 425 galaxies, of which 401 are early-types.  The galaxies were selected from
the Lyon-Meudon Extragalactic Data Archive (LEDA) to satisfy the following criteria.
\begin{enumerate}
\item Morphological type $T<-1.5$ (i.e.\ E and S0 galaxies).
\item Virgo-corrected recession velocity $V\le 9000$km s$^{-1}$.
\item Apparent magnitude $B_{T}\le 13.5$.
\end{enumerate}

Since LEDA is $\approx 90\%$ complete at $B_{T}=14.5$, criteria (ii) and (iii) should provide a sample with high
statistical completeness.  This selection yielded $\sim 700$ galaxies.  For 136 of these, \citet{osul01} used PSPC data from archival
ROSAT pointings to derive X-ray luminosities. Circular extraction regions
were chosen, with cut-off radii set at the point where the source emission
dropped to the background level. In the case of groups and clusters, the
surrounding intra-cluster medium was considered as the background. Spectra
were extracted and fit with a fixed 1 keV, solar abundance MEKAL model
(\citealt{kaa93}; \citealt{lie95}). Hydrogen absorption column
densities were fixed at galactic values using the survey of \citet{sta92}. Such a model is clearly an over-simplification of the actual
spectrum of emission from early-type galaxies, but was chosen to allow
easy comparison with other catalogues which use similar models, to be
relatively representative of the majority of early-type galaxies, and to
allow fitting of a simple model to the entire dataset. While more
complicated multi-component models would have provided more accurate
estimates of the properties of the best-observed systems, they would not
have been applicable to those systems with few counts, and would have made
direct comparisons difficult. Bolometric X-ray luminosities were
determined, and the authors note that if the lowest luminosity galaxies
were in fact better fitted by a 10 keV bremsstrahlung model (an
approximation of the spectrum expected from a galaxy dominated by X-ray
binaries), these luminosities would be underestimated by only a factor of
$\sim$2.

For another 265 galaxies, X-ray luminosities were available from the
surveys of \citet{beu99}, \citealt{fab92} or \citet{rob91}. These surveys were based on ROSAT All-Sky Survey data or on
Einstein Imaging Proportional Counter pointed data, and made use of a range of spectral models,
wavebands and distance scales. Corrections were therefore made to bring the
X-ray luminosities from these catalogues into line with those of
\citet{osul01}, and tests performed using galaxies listed in
more than one sample, to ensure that the final luminosities were
comparable. A small number of objects with divergent luminosities were
excluded, but these were mainly found to be AGN, or confused with other
sources. After correction, the X-ray luminosities from the literature were
found to be in good agreement with those derived by O'Sullivan et al.,
producing a final sample of 401 galaxies (of which $\sim$50\% have only
X-ray upper limits) , the most extensive catalogue of its type to date.
Further details of this sample and its construction can be found in
\citet{osul01}. The size of the catalogue is its main strength
for our purposes; searching for correlations with other properties
requires a large sample in order to ensure statistical accuracy,
particularly when working with upper limits. The main weaknesses of the
catalogue are the inherent inaccuracies associated with the estimation of
the \lx\ values. Both the choice of model and the definition of
extraction region could potentially lead to errors in the final \lx.
However, in the vast majority of cases, neither factor should cause severe
inaccuracies. It should also be noted that improving on the sample is not
currently possible; other samples of the ROSAT/Einstein era are smaller
and in some case less well defined, and while the \emph{Chandra} and
\emph{XMM Newton} observatories could provide the collecting area,
spectral and spatial resolution required for a superior catalogue, the
number of suitable ellipticals available in their archives is as yet too
small for our purpose.


\section{Near-infrared data}
\label{sec:nir}

Near-infrared photometry has several advantages over optical photometry as an indicator of galaxy mass.  The flux in
the NIR is dominated by old, long-lived stellar populations, whereas blue luminosities can be greatly enhanced by a
relatively small number of young, bright, massive stars.  Furthermore the galactic extinction in the NIR is much
smaller than at optical wavelengths, reducing the uncertainty introduced in correcting for extinction.

The 2MASS covers 99.998\% of the sky in the \emph{J}, \emph{H} and \emph{Ks} bands.  Sensitivity for extended sources is $\sim$ 14.7, 13.9 and
13.1 mag at \emph{J}, \emph{H} and \emph{Ks} respectively (\citealt{jar00}).  \emph{H} and \emph{Ks} band magnitudes have been extracted from the
Extended Source Catalogue (XSC, \citealt{jar00}) and the Large Galaxy Atlas (LGA, \citealt{jar03}) using the NASA
Extragalactic Database (NED).  Where magnitudes were available from both catalogues the LGA was preferred since it
accounts for the incompleteness of the XSC for galaxies greater than $\sim 2'$ diameter, which may be incorrectly sampled if
they are located close to the edge of a 2MASS scan, since they are larger than the overlap between scans, and
consequently may be truncated.  Thus the
LGA constructs mosaics of the largest galaxies and derives magnitudes from these.  Throughout we have used
\citet{kro80} magnitudes as reported from NED queries.

NIR photometry has been obtained for 400 of the 425 galaxies catalogued by \citet{osul01}, of which 376 are
early-types.  Magnitudes have
been converted to units of \emph{H} and \emph{Ks} solar luminosities assuming $M_{{\rm V\odot}}=4.82$ (\citealt{allens00}),
$V-H_{\odot}=1.44$
(\citealt{colina97}) and $V-K_{s\odot}=1.45$ (\citealt{tof03}).  Note that while the conversion to units of solar
luminosity may not be absolutely calibrated for the 2MASS filters used (although they should not be very different
from the values quoted), any error in the value of $K_{s\odot}$ and
$H_{\odot}$ will only result in an error in the normalisation of the \lx:\lk\ and \lx:\lh\ relationships.  The slope
and scatter will be unaffected.  Since we are
mainly concerned with the slope and scatter of the relation, absolute calibration of the magnitude of the sun in the
2MASS filters is
not  needed.

The X-ray, NIR and optical luminosities are listed, along with other data used in this paper, in Table~\ref{tab:bigtab}.

\section{Statistical analysis}
\label{sec:method}

The fitting of \lx:\lb, \lx:\lh\ and \lx:\lk\ relations is complicated by the fact that many of the X-ray detections
are upper limits.  To deal with this the method of \citet{osul01} has been repeated for the NIR relations, and is
summarised here.

Survival analysis has been used to estimate the true values associated with the upper limits. Survival analysis assumes
that the upper limits are unrelated to the true values, and uses the distribution of detections in the catalogue below
each upper limit to model the probability distribution of the true value associated with that upper limit. 
\citet{osul01} show that this is a reasonable assumption for the catalogue since for the majority of the galaxies
there is no relation between the luminosity of the source and the sensitivity of the observation due to the large
number of serendipitous detections.  See \citet{osul01} for a full discussion of potential sources of bias in the sample and the benefits and weaknesses of the fitting methods used.

To fit lines to the relations two methods have been used: the expectation and maximisation algorithm (EM) and the
Buckley-James algorithm (BJ)\footnote{See {\sc iraf} task stsdas.analysis.statistics.survival for an introduction to survival analysis.} using the {\sc asurv} software package (\citealt{lav92}).  The EM method is parametric and assumes the residuals to the line follow a Gaussian
distribution.  The BJ method is non-parametric and calculates the regression using the \citet{kap58} estimator for the
residuals (see \citealt{fei85}).  In principal the two methods should agree, disagreement would suggest the fits are not  robust.  We find very good agreement between the two methods in all cases, and hence we only show the EM fits in the plots throughout the paper.

\section{Results}
\label{sec:results}

\begin{figure*}
\centering \includegraphics[scale=0.7,angle=0]{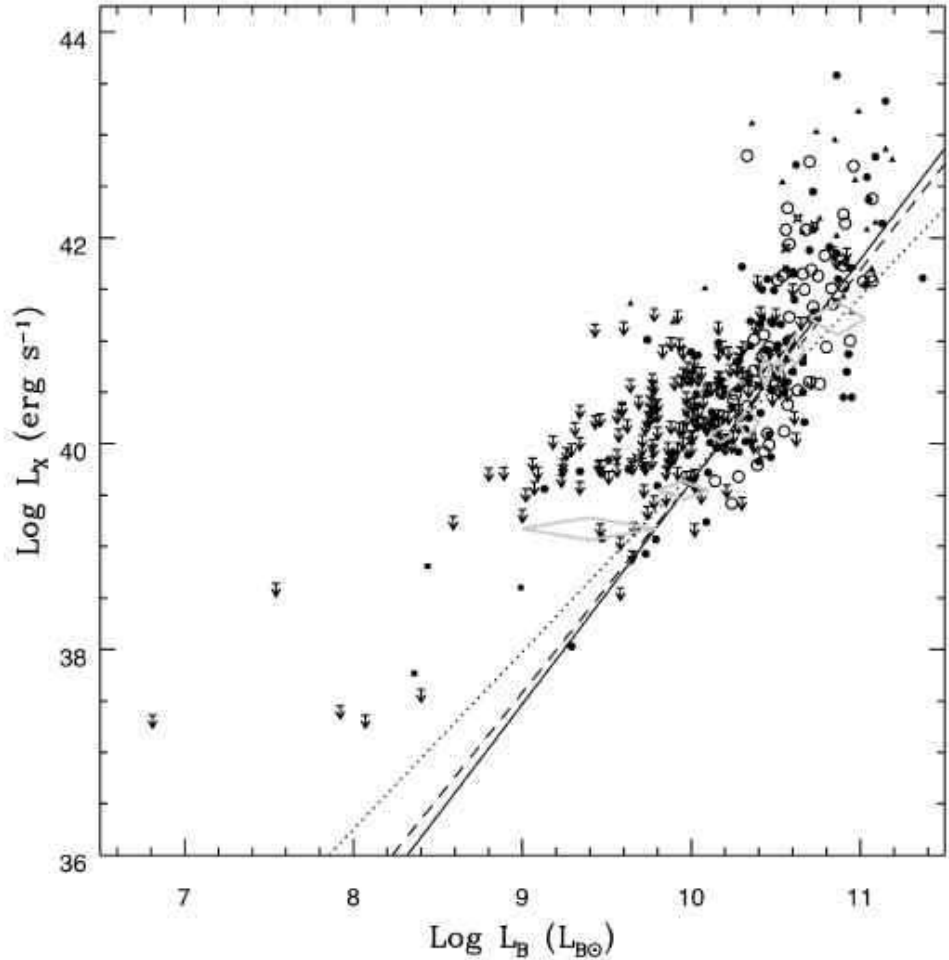}
\caption{log\lx\ as a function of log\lb\ for all early-type galaxies.  Triangles represent BCGs, open circles are BGGs, stars are AGN, arrows are upper limits of any type.  The lines
show the best fit using the EM algorithm.  Line styles indicate the sample used in the
fits as discussed in the text; solid uses subsample (i), dashed lines uses subsample (ii), dotted uses subsample (iii).  The diamonds show the average \lx\ in bins
containing at least 10 detections.  Optically bright galaxies display a steeper relation than faint galaxies.}
\label{fig:lxlb}
\end{figure*}

\begin{figure*}
\centering \includegraphics[scale=0.7,angle=0]{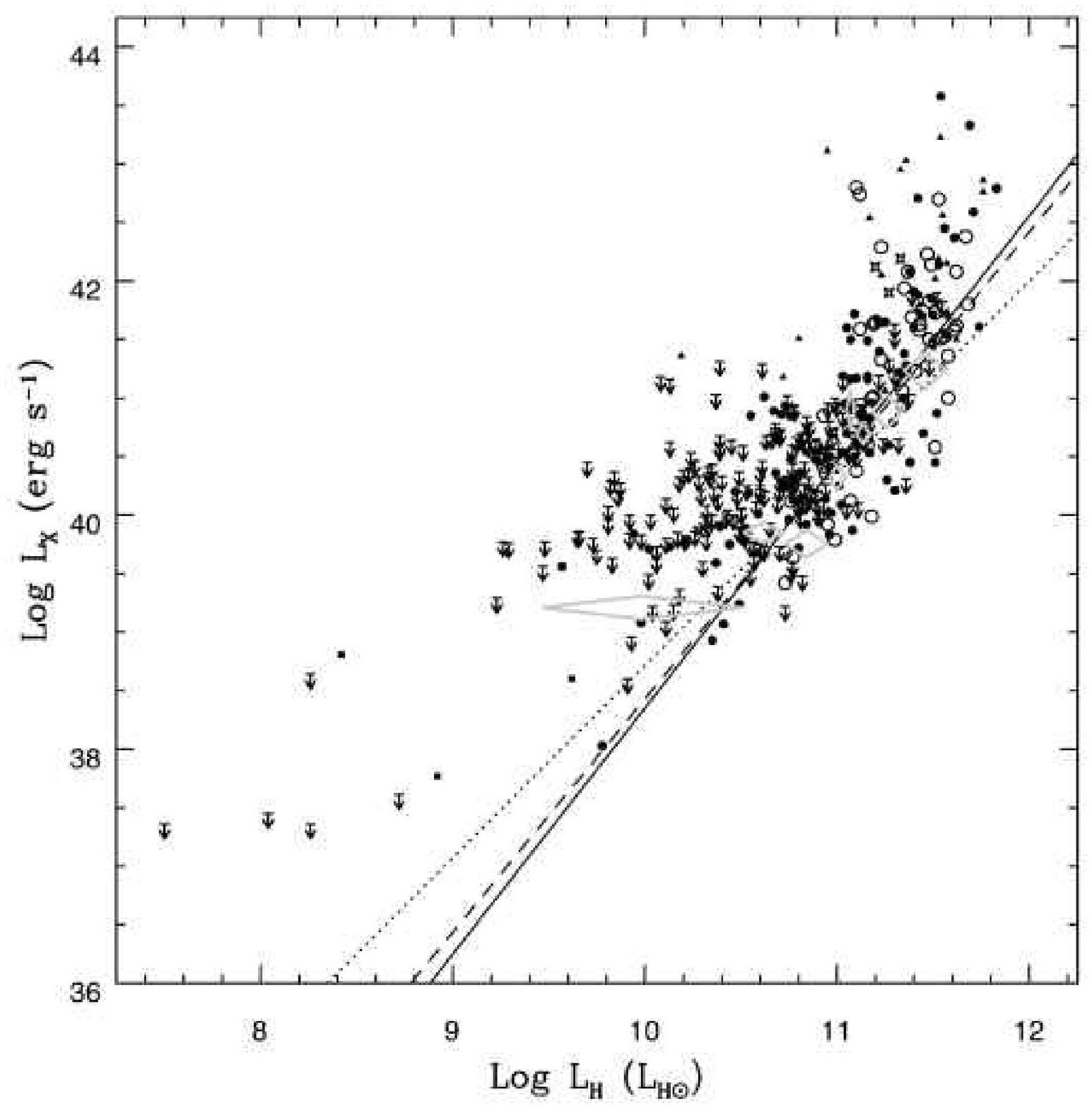}
\caption{log\lx\ as a function of log\lh.  Symbols are as for Figure~\ref{fig:lxlb}.  The log\lh\ relation is consistent with the log\lb\ relation.}
\label{fig:lxlh}
\end{figure*}

\begin{figure*}
\centering \includegraphics[scale=0.7,angle=0]{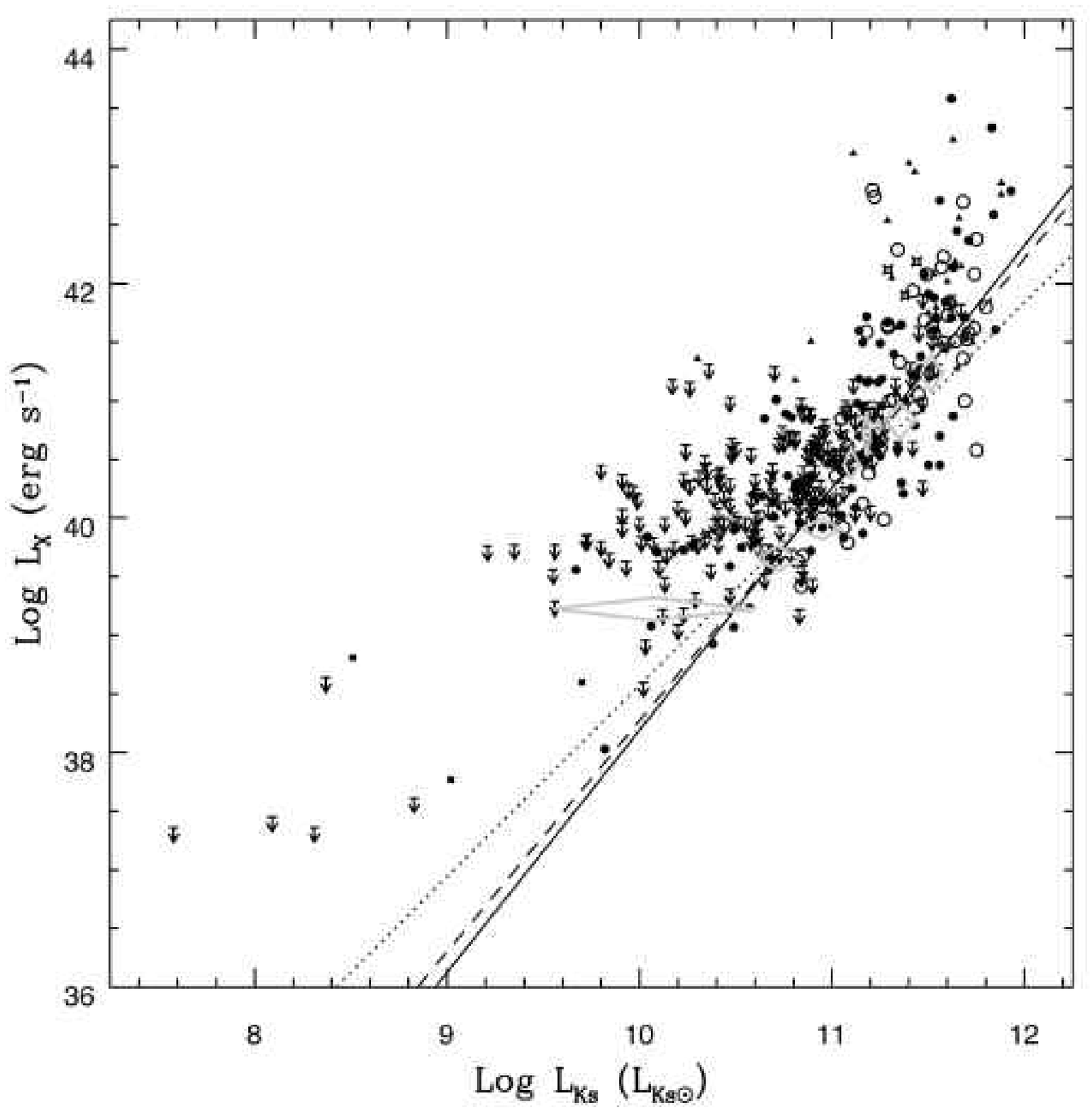}
\caption{log\lx\ as a function of log\lk.  Symbols are as for Figure~\ref{fig:lxlb}.  The log\lk\ relation is consistent with the log\lb\ relation.}
\label{fig:lxlk}
\end{figure*}

The relation between \lx\ and \lb, \lh, \lk\ has been investigated for all early-type galaxies for which NIR data have
been obtained.  
Plots of log\lx\ vs.\ log\lb, log\lh\ and log\lk\ are shown in Figures~\ref{fig:lxlb}, \ref{fig:lxlh} and \ref{fig:lxlk}
respectively.  Galaxies were sub-divided into brightest cluster galaxies (BCGs),  brightest group galaxies (BGGs), AGN
and dwarfs (log (\lb/\lbsun) $<9$).  Fits have been made using three groups.
\begin{enumerate}
\item All galaxies excluding late-types, AGN, BCGs and dwarfs (solid line).
\item As (i) but also excluding BGGs (dashed line).
\item As (ii) but also excluding galaxies at a distance $>70$ Mpc (for which environment is not robust), NGC 5102
 and NGC 4782  (dotted line).  NGC 5102 (which has very low \lx/\lb) and NGC 4782 (which has very high \lb) drive the fits unless
excluded (see \citealt{osul01} for a discussion of the unusual \lx/\lb\ values for these galaxies).
\end{enumerate}

The best fits for each subsample are listed in Table~\ref{tab:fit} showing the strong agreement between colours.  The different subsamples are all in good agreement with the exception of the BGGs, which show a steeper relation, in agreement with previous work (\citealt{osul01}; \citealt{hel01}).
All three plots are remarkably similar.  Each shows the characteristic transition to steeper slopes for optically brighter (more
massive)
galaxies, as the hot gas contribution to the X-ray luminosity becomes more significant.  This is apparent in
comparing the average values of \lx\ as denoted by the grey diamonds to the best fit lines, which match well at high
\lb, \lh, \lk, but not at lower values where the contribution to \lx\ from discrete sources dominates.  The scatter
appears similar in each band, with the \emph{H}-band
possibly giving the best fits, since the agreement between the different fitting algorithms is strongest.  

\begin{table*}
\begin{center}
\begin{tabular}{lccccccc}
\hline \\
Sample & N$_{det}$ & N$_{UL}$ & Band & Method & Slope (error) & Intercept (error) & Std. Dev.\\
\hline
\hline
Combined (i) & 155 & 173 & $B$ & EM & 2.16 ($\pm$0.12) & 18.03 ($\pm$1.25) & 0.69\\
 & & & $B$ & BJ & 2.19 ($\pm$0.14) & 17.74 & 0.67\\
 & & & $H$ & EM & 2.10 ($\pm$0.11) & 17.33 ($\pm$1.20) & 0.64\\
 & & & $H$ & BJ & 2.09 ($\pm$0.12) & 17.46 & 0.61\\
 & & & $K_S$ & EM & 2.07 ($\pm$0.11) & 17.53 ($\pm$1.19) & 0.63\\
 & & & $K_S$ & BJ & 2.21 ($\pm$0.12) & 15.90 & 0.63\\
\hline
Combined (ii) & 99 & 153 & $B$ & EM & 2.05 ($\pm$0.14) & 19.10 ($\pm$1.41) & 0.69\\
 & & & $B$ & BJ & 2.05 ($\pm$0.15) & 19.17 & 0.67\\
 & & & $H$ & EM & 1.99 ($\pm$0.12) & 18.48 ($\pm$1.36) & 0.64\\
 & & & $H$ & BJ & 1.99 ($\pm$0.14) & 18.58 & 0.60\\
 & & & $K_S$ & EM & 1.98 ($\pm$0.12) & 18.55 ($\pm$1.33) & 0.63\\
 & & & $K_S$ & BJ & 1.97 ($\pm$0.13) & 18.65 & 0.59\\
\hline
Combined (iii) & 83 & 153 & $B$ & EM & 1.72 ($\pm$0.14) & 22.47 ($\pm$1.48) & 0.60\\
 & & & $B$ & BJ & 1.71 ($\pm$0.16) & 22.60 & 0.56\\
 & & & $H$ & EM & 1.64 ($\pm$0.13) & 22.28 ($\pm$1.38) & 0.56\\
 & & & $H$ & BJ & 1.63 ($\pm$0.14) & 22.41 & 0.51\\
 & & & $K_S$ & EM & 1.63 ($\pm$0.12) & 22.23 ($\pm$1.37) & 0.55\\
 & & & $K_S$ & BJ & 1.76 ($\pm$0.14) & 20.83 & 0.52\\
\hline
Cluster & 20 & 35 & $B$ & EM & 1.80 ($\pm$0.28) & 21.68 ($\pm$2.80) & 0.57\\
 & & & $B$ & BJ & 1.79 ($\pm$0.30) & 21.73 & 0.52\\
 & & & $H$ & EM & 1.55 ($\pm$0.24) & 23.21 ($\pm$2.55) & 0.56\\
 & & & $H$ & BJ & 1.54 ($\pm$0.25) & 23.41 & 0.49\\
 & & & $K_S$ & EM & 1.55 ($\pm$0.23) & 23.13 ($\pm$2.53) & 0.56\\
 & & & $K_S$ & BJ & 1.53 ($\pm$0.25) & 23.34 & 0.48\\
\hline
Field & 21 & 52 & $B$ & EM & 1.61 ($\pm$0.23) & 23.64 ($\pm$2.40) & 0.68\\
 & & & $B$ & BJ & 1.60 ($\pm$0.26) & 23.71 & 0.70\\
 & & & $H$ & EM & 1.67 ($\pm$0.20) &  21.89 ($\pm$2.26) & 0.59\\
 & & & $H$ & BJ & 1.67 ($\pm$0.22) & 21.89 & 0.58\\
 & & & $K_S$ & EM & 1.70 ($\pm$0.21) & 21.71 ($\pm$2.31) & 0.59\\
 & & & $K_S$ & BJ & 1.70 ($\pm$0.23) & 21.70 & 0.59\\
\hline
Group (total) & 96 & 82 & $B$ & EM & 2.20 ($\pm$0.19) & 17.64 ($\pm$1.95) & 0.64\\
 & & & $B$ & BJ & 2.18 ($\pm$0.22) & 17.78 & 0.62\\
 & & & $H$ & EM & 2.10 ($\pm$0.16) & 17.34 ($\pm$1.82) & 0.60\\
 & & & $H$ & BJ & 2.09 ($\pm$0.19) & 17.51 & 0.57\\
 & & & $K_S$ & EM & 2.06 ($\pm$0.16) & 17.59 ($\pm$1.80) & 0.60\\
 & & & $K_S$ & BJ & 2.04 ($\pm$0.19) & 17.82 & 0.57\\
\hline
Group (non-BGG) & 44 & 66 & $B$ & EM & 1.75 ($\pm$0.22) & 22.22 ($\pm$2.29) & 0.55\\
 & & & $B$ & BJ & 1.73 ($\pm$0.26) & 22.46 & 0.59\\
 & & & $H$ & EM & 1.67 ($\pm$0.20) & 22.04 ($\pm$2.16) & 0.51\\
 & & & $H$ & BJ & 1.65 ($\pm$0.23) & 22.23 & 0.48\\
 & & & $K_S$ & EM & 1.67 ($\pm$0.20) & 21.86 ($\pm$2.16) & 0.51\\
 & & & $K_S$ & BJ & 1.65 ($\pm$0.23) & 22.05 & 0.48\\
\hline
Group (BGG only) & 53 & 16 & $B$ & EM & 2.58 ($\pm$0.36) & 13.60 ($\pm$3.76) & 0.69\\
 & & & $B$ & BJ & 2.57 ($\pm$0.40) & 13.71 & 0.68\\
 & & & $H$ & EM & 2.47 ($\pm$0.30) & 13.23 ($\pm$3.32) & 0.63\\
 & & & $H$ & BJ & 2.43 ($\pm$0.32) & 13.69 & 0.62\\
 & & & $K_S$ & EM & 2.35 ($\pm$0.29) & 14.32 ($\pm$3.24) & 0.64\\
 & & & $K_S$ & BJ & 2.31 ($\pm$0.32) & 14.87 & 0.63\\
 \hline
\end{tabular}
\end{center}
\caption{Best fit lines for the various correlations discussed in the text. Columns 2 and 
3 show the number of detections and upper limits in each sample, respectively. Column 8 shows the standard deviation of the residual in log L$_X$ about each fit line, providing an 
estimate of the degree of scatter.}
\label{tab:fit}
\end{table*}

\subsection{Environment}

\begin{figure*}
\centering \includegraphics[scale=0.9,angle=0]{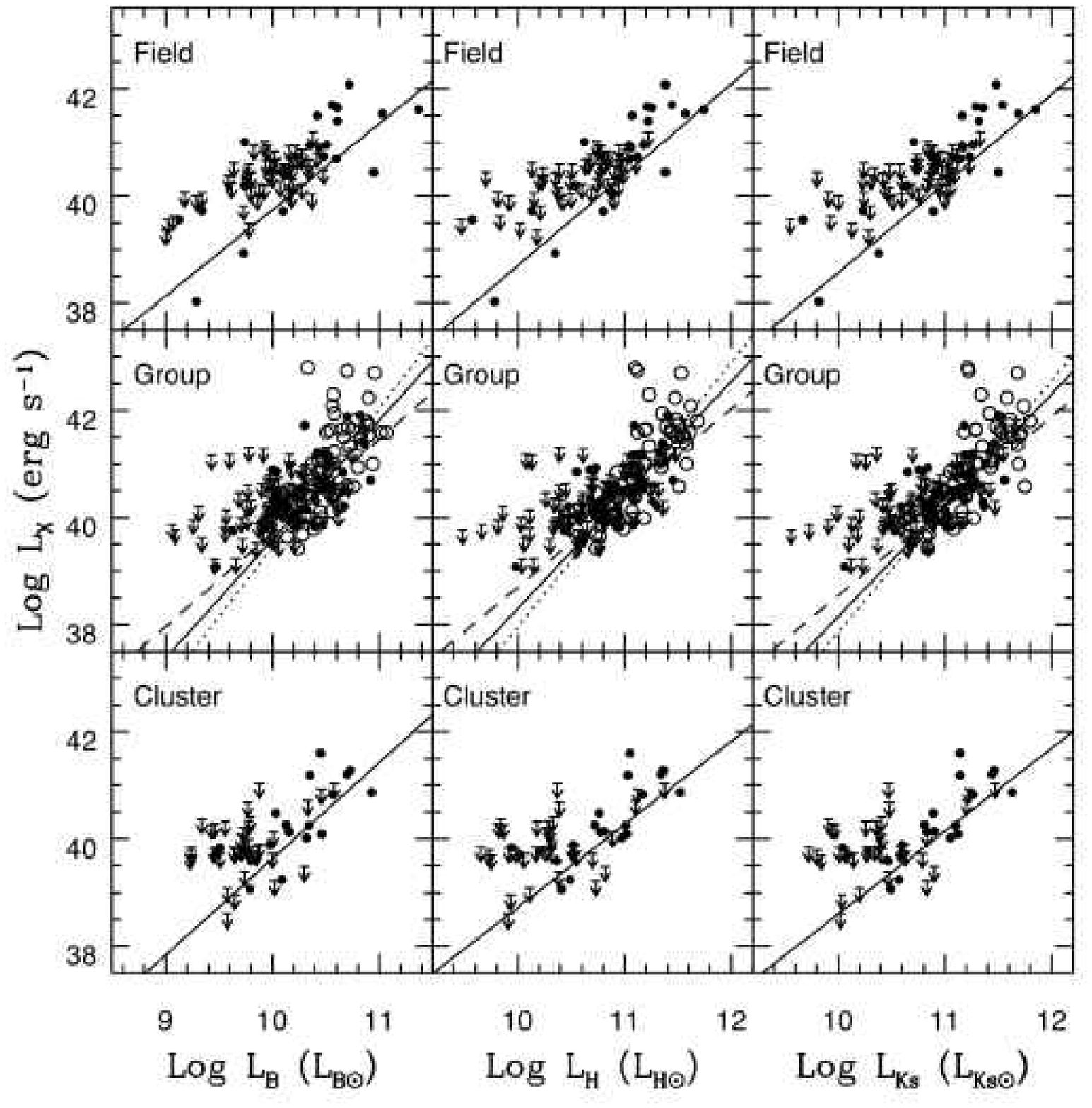}
\caption{log \lx\ as a function of log\lb, log\lh\ and log\lk.  The top panel shows field galaxies, the middle panel shows group galaxies and the bottom panel shows cluster galaxies.  All AGN, dwarfs, late-types, BCGs and galaxies with a
distance $>70$ Mpc have been excluded.  Filled circles are detections, open circles are BGGs, arrows are upper limits.  The lines show the best fits using the EM algorithm.  Solid lines are for the whole sample, dashed lines exclude BGGs and dotted lines are for BGGs only.  There is little difference with waveband or density with the exception of BGGs, which display a steeper relation.}
\label{fig:env}
\end{figure*}

The correlations between luminosities have been repeated for galaxies segregated according to environment. 
Figure~\ref{fig:env} shows plots for \lb, \lh\ and \lk\ respectively split
into field, group and cluster galaxies.  Cluster membership has been determined from \citet{abe89} and \citet{faber89}
and
group membership has been determined from \citet{gar93}.  The sample is limited to galaxies within $V=5500$km s$^{-1}$
by the \citet{gar93} catalogue.  Comparison of the panels in Figure~\ref{fig:env} again reveals consistency between the different 
wavebands.

The large scatter around the \lx:\lb\ relation has often been speculated to be a consequence of environment (\citealt{mat03}). \lx\ could be
reduced in high density environments due to mechanisms such as ram-pressure stripping of the gaseous haloes.  Alternatively an
enhancement of \lx\ is possible  due to the stifling of galactic winds by the surrounding intragroup or intracluster gas, and an increase of
the gas density due to the surrounding ICM, or due to accretion of gas and/or dark matter from the surrounding medium.  The evidence for such effects is inconclusive.  Some galaxies located in the central
regions of groups or
clusters have enhanced \lx\ due to a contribution from the intragroup or intracluster medium (\citealt{hel01}; \citealt{mat01}). 
On the other hand, \citet{whit91} and \citet{hen99} find that galaxies have lower \lx/\lb\ if the local galaxy density is high.  However
\citet{bro00} find the opposite effect with galaxies in high density regions having higher \lx/\lb.   \citet{osul01}, in their larger sample, find no trend with environment except for BGGs, which have larger \lx/\lb, which is thought to drive the result of \citet{bro00} due to the large fraction of BGGs in their sample (\citealt{hel01}).

\begin{figure}
\begin{minipage}[c]{0.3\textwidth}
\centering \includegraphics[scale=0.39,angle=0]{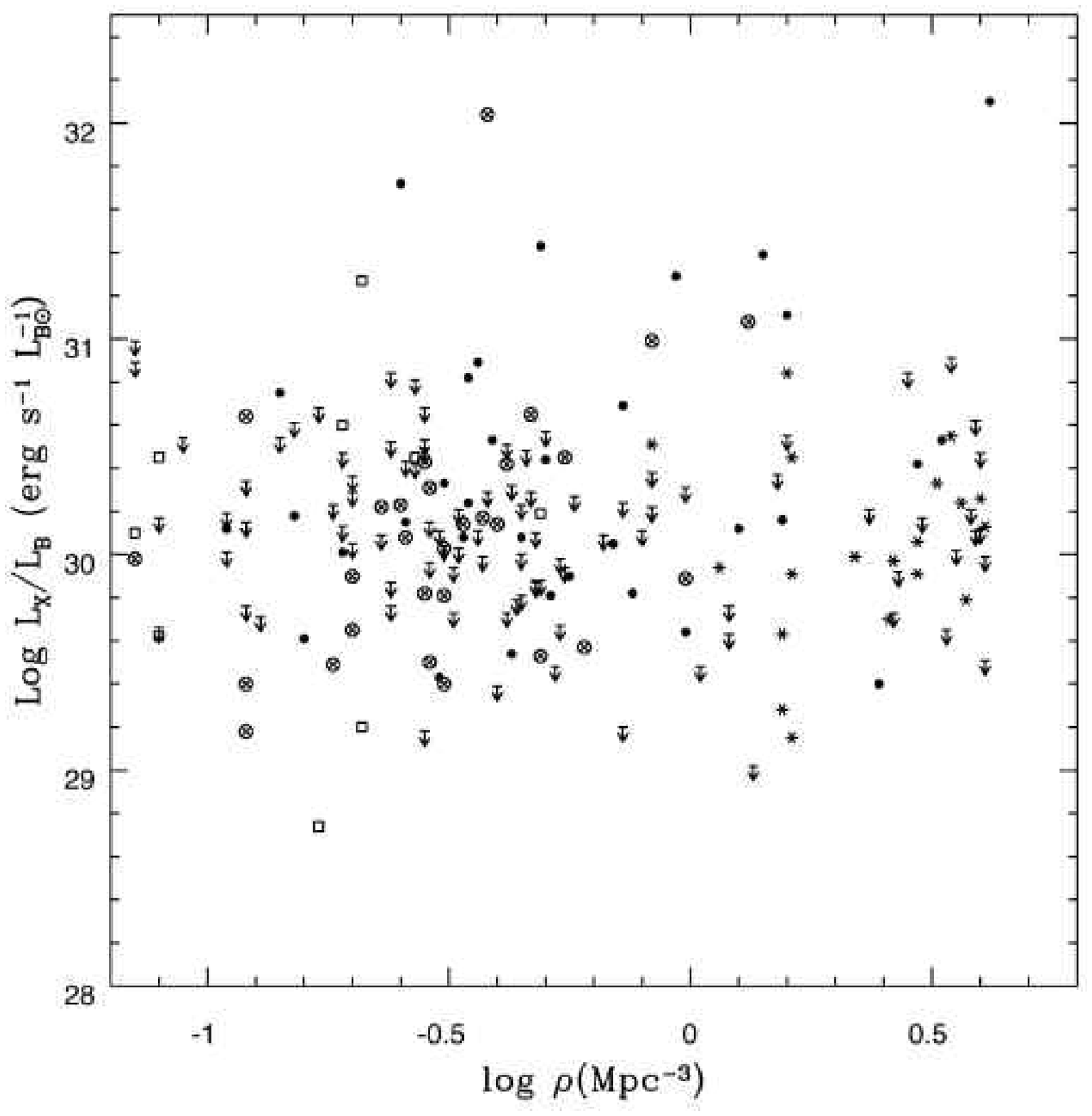}
\end{minipage}
\begin{minipage}[c]{0.3\textwidth}
\centering \includegraphics[scale=0.39,angle=0]{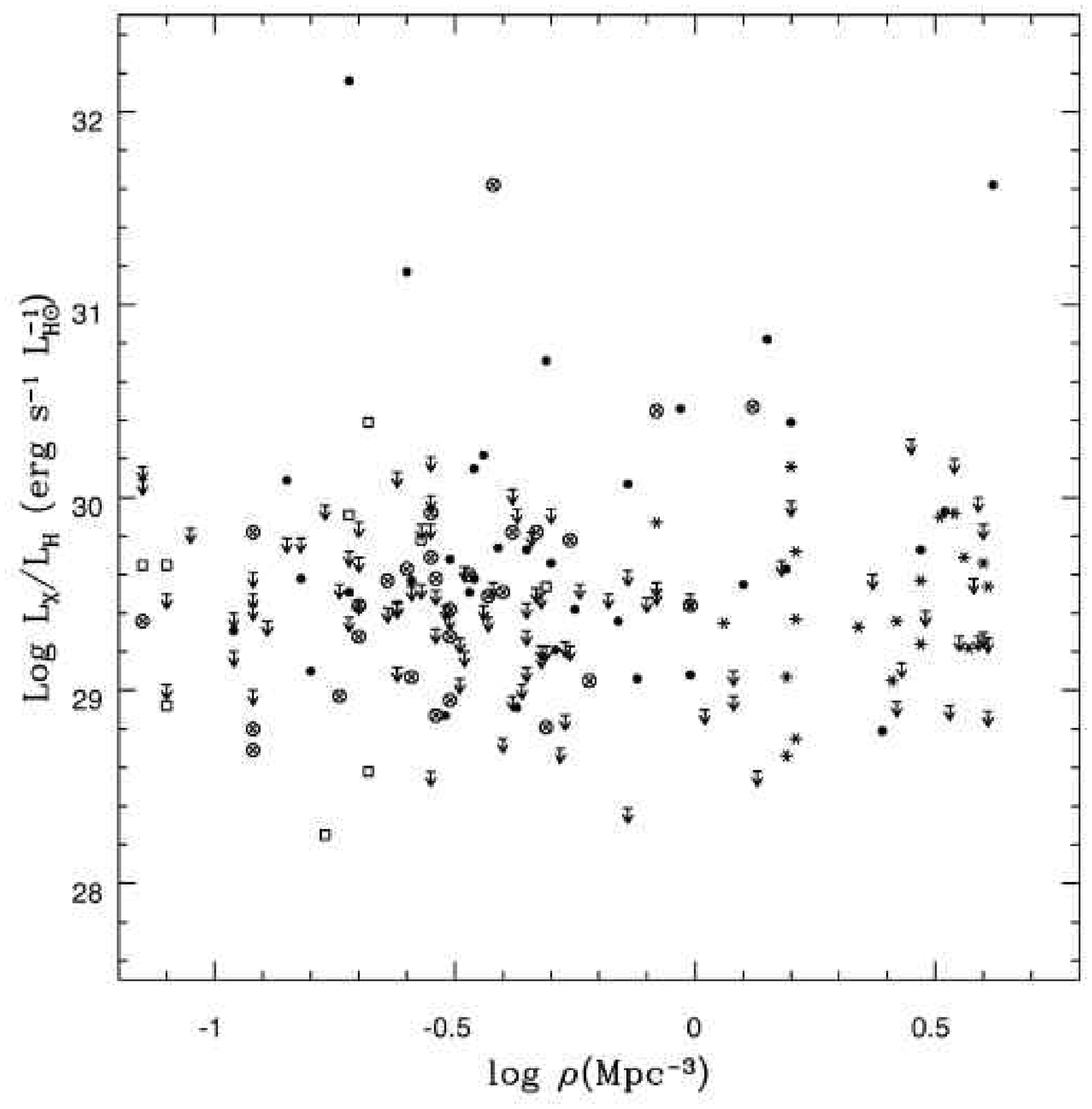}
\end{minipage}
\begin{minipage}[c]{0.3\textwidth}
\centering \includegraphics[scale=0.39,angle=0]{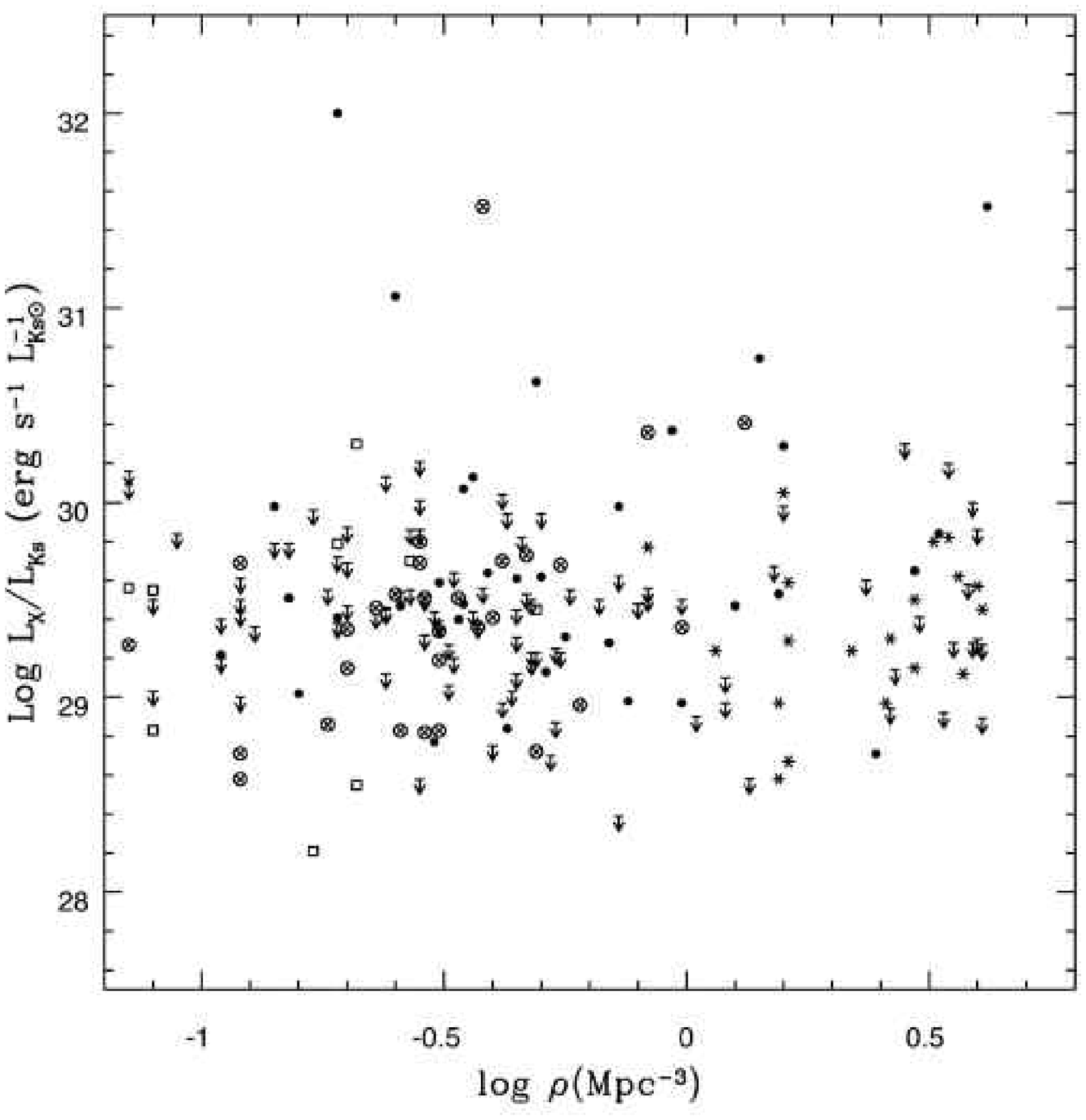}
\end{minipage}  
\caption{Ratios of \lx\ and optical/NIR luminosities as a function of \citet{tul88} density.  Asterisks represent cluster galaxies, crossed circles
represent BCGs, open circles represent group galaxies, squares represent field galaxies.  There is no trend with density and the scatter is similar in all wavebands.}
\label{fig:tully}
\end{figure}

The effect of environment has been re-examined in the NIR bandpasses.  The scatter around the relations of Figure~\ref{fig:tully}
has been calculated using the Kaplan-Meier estimate of the means, and are listed in
Table~\ref{tab:scatter} for different environment and waveband.  For samples in which the lowest point is an upper-limit, the results may be biased
low, since the Kaplan-Meier estimate assumes that the point is a detection.  Where this is the case the lower limits have also been given.

\begin{table}
\caption{The means and scatters of the relations of log\lx/\lb, log\lx/\lh\ and
log\lx/\lk\ for galaxies located in different environments.} 
\begin{tabular}{lccc}
\hline \\
Relation & Mean & Error on mean & Limit \\ \hline
& \multicolumn{3}{c}{{\bf Field}} \\
\lx/\lb & 29.783 & 0.144 &---\\
\lx/\lh & 29.073 & 0.127 &---\\
\lx/\lk & 29.146 & 0.136 &---\\
& \multicolumn{3}{c}{{\bf Clusters}} \\
\lx/\lb & 29.710 & 0.097 & 29.020 \\
\lx/\lh & 29.023 & 0.093 & 28.390 \\
\lx/\lk & 29.111 & 0.091 & 28.490 \\
& \multicolumn{3}{c}{{\bf Groups - total}} \\
\lx/\lb & 30.069 & 0.052 &---\\
\lx/\lh & 29.365 & 0.049 &---\\
\lx/\lk & 29.461 & 0.049 &---\\
& \multicolumn{3}{c}{{\bf Groups - excluding BGGs}} \\
\lx/\lb & 29.936 & 0.055 & 29.390 \\
\lx/\lh & 29.232 & 0.053 & 28.700 \\
\lx/\lk & 29.321 & 0.053 & 28.780 \\
& \multicolumn{3}{c}{{\bf Groups - BGGs only}} \\
\lx/\lb & 30.298 & 0.091 &---\\
\lx/\lh & 29.576 & 0.086 &---\\
\lx/\lk & 29.677 & 0.086 &---\\
\hline
\end{tabular}
\label{tab:scatter}
\end{table}

The scatter in the NIR bands is only marginally smaller than the \emph{B}-band scatter, implying that the variance in X-ray emission dominates the
scatter.  This is expected if ellipticals are largely composed of old, quiescent stellar populations possessing relatively little dust.  In
such a case ellipticals will be optically thin, and both  NIR and optical light will be dominated by emission from old, long-lived stars.
 
There is no clear trend in the mean luminosity ratios with environment.  Means are similar throughout all environments examined, with the exception of BGGs.  The extremes
of the environments investigated, viz.\ field galaxies and cluster galaxies, are consistent.  However, the group environment has slightly higher
average values in all bands, and the BGGs have the highest ratios of all.

The high X-ray emission of BGGs may be due to a unique formation mechanism (\citealt{bha85}) or may be indicative that processes such as accretion or stifling of galactic winds are active in the group environment.  The slightly higher X-ray emission of the group galaxies in general, if real, is supportive of the idea that IGM influences the X-ray emission of galaxies in groups, but the enhancement is stronger for BGGs due their typical location at the centre of the group potential well, where the
IGM is densest.  However, no such enhancement is seen for cluster galaxies, which would be expected to have a similar enhancement due to the ICM, unless some counter-acting process is responsible such as ram-pressure stripping, which may only be effective in the denser environments of clusters (\citealt{fuj01}).

A further test of the influence of environment on the scatter is illustrated in Figure~\ref{fig:tully}.  The ratios of log\lx/\lb, log\lx/\lh\
and log\lx/\lk\ are plotted as a function of \citet{tul88} density parameter, for all galaxies in the Tully Nearby Galaxy Catalogue.  There is no clear
relation in any of the luminosity ratios with local galaxy density, and the scatter appears similar in all wavebands.

As a caveat, we note that this sample does not contain any very high density environments such as rich clusters.  Recent \emph{Chandra} observations presented by \citet{sun05} show that in the dense environment of A1367 it is likely that only the most massive galaxies are able to retain their X-ray haloes.    The haloes of less massive galaxies are probably stripped through interaction with the dense surrounding medium.

\subsection{Scatter as a function of core profile}

\citet{pel99} shows that the global X-ray emission of early-type galaxies is correlated with
their
internal central properties.  Specifically `core' galaxies, defined as having an inner slope of $\gamma <
0.3$, where $I\propto R^{-\gamma}$, show a large scatter in values of \lx\ whereas `power-law' galaxies
with $\gamma > 0.5$ are restricted to \lx\ $< 10^{41}$ erg s$^{-1}$.  Furthermore core galaxies are
associated with boxy, slowly rotating, optically luminous galaxies, whereas power-law galaxies are discy, rapidly rotating systems, which are generally fainter.

It is unlikely that the central stellar components of early-type galaxies are directly influencing the surrounding X-ray haloes.  Rather the link between $\gamma$ and \lx\ is more likely to be indirect, since both properties are known to be correlated with \lb.  However, since both \lx\ and $\gamma$ could also be linked with formation processes, the form of the relation between the two properties is important.

This correlation has been investigated here, taking determinations of core/power-law and values of $\gamma$ and $a(4)/a$ from the work of \citet{pel99}, \citet{rest01}, \citet{rav02}, \citet{mic98}, \citet{fab97} and \citet{ben89}.  For galaxies appearing in more than one paper, preference was in the order of the papers as listed, with the following exceptions.  NGC3377 and NGC4564 appear in \citet{pel99} as power-law galaxies with $\gamma<0.3$, so updated values from \citet{rest01} are preferred.   Similarly for the core galaxy NGC4374, \citet{pel99} quote a value of $\gamma=0.31$, which has been replaced by the value of $\gamma=0.13$ from \citet{rav02}.  In total there are 54 power-law galaxies, 30 core galaxies, 74 galaxies have measured $\gamma$ and 93 galaxies have measured $a(4)/a$.  

\begin{figure}
\centering \includegraphics[scale=0.4,angle=0]{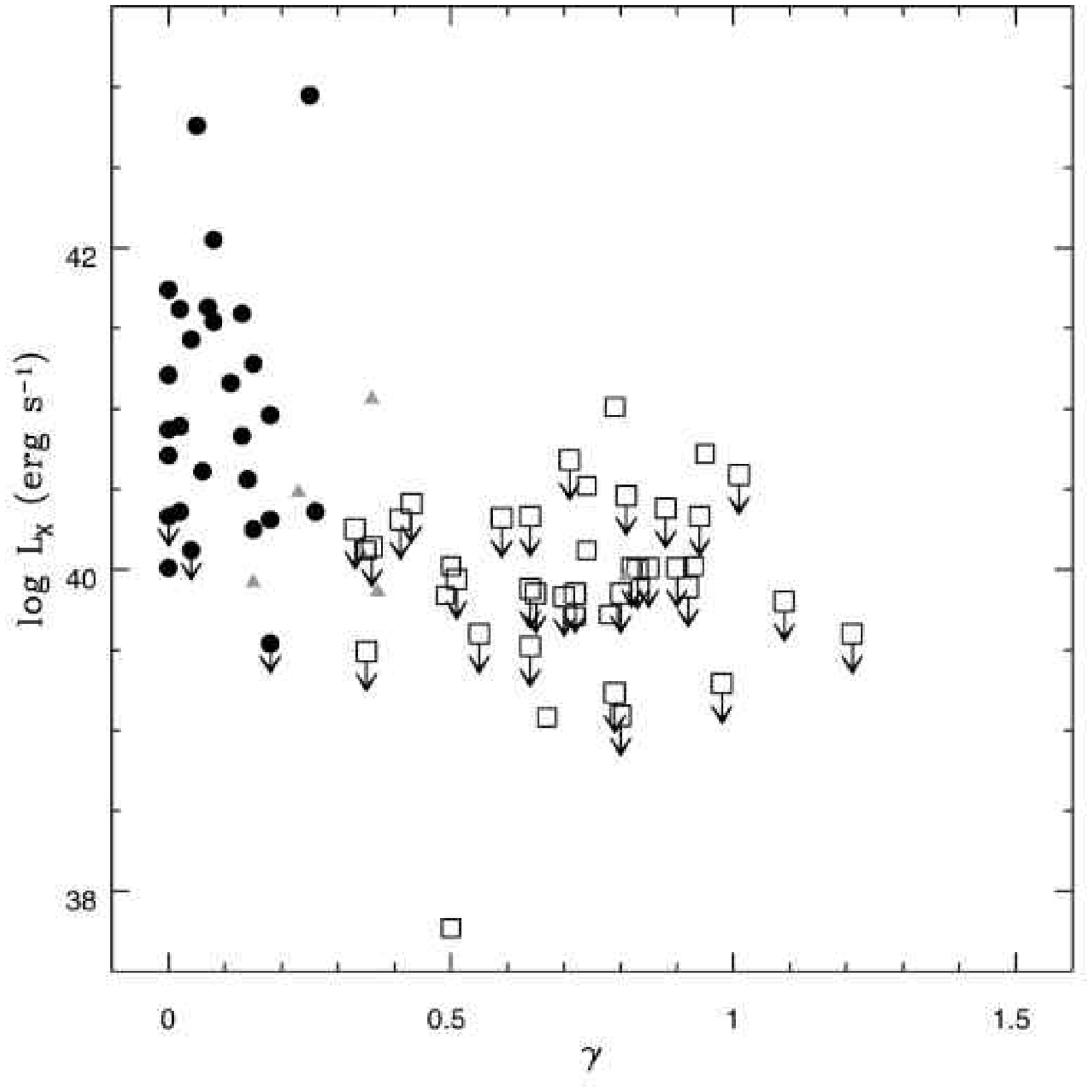}
\caption{log\lx\ as a function of $\gamma$.  Circles indicate core galaxies while squares indicate power law galaxies, triangles are for undetermined central profiles.  Core galaxies have a larger range in \lx\ than power-law galaxies, which are generally confined to log\lx$<41$ erg s$^{-1}$.  The outlier at log\lx$<38$ erg s$^{-1}$ is M32. }
\label{fig:lxgamma}
\end{figure}

\begin{figure}
\centering \includegraphics[scale=0.4,angle=0]{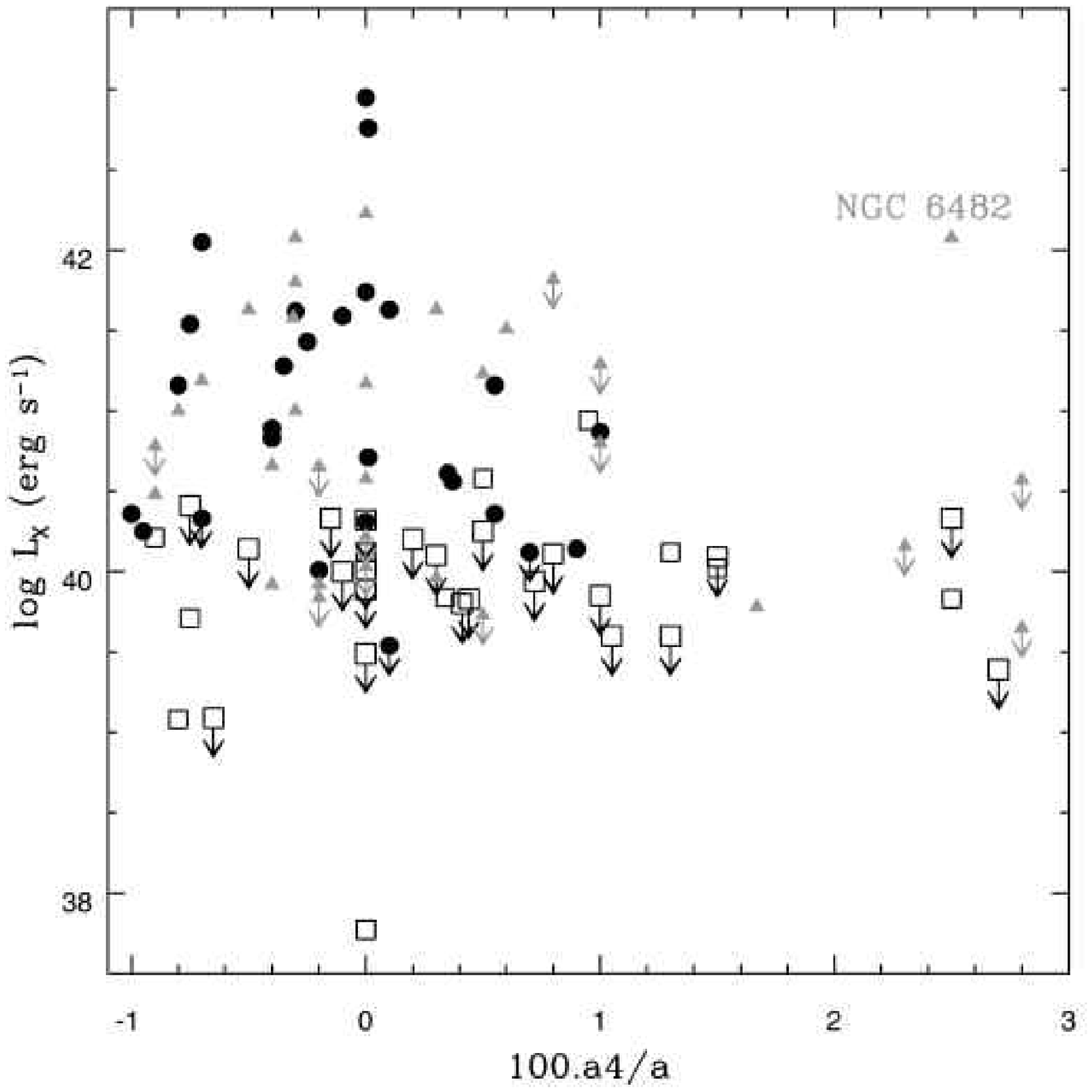}
\caption{log\lx\ as a function of $100 \times a(4)/a$.  Circles indicate core galaxies while squares indicate power law galaxies, triangles are for undetermined central profiles.  Galaxies which deviate from a pure ellipse are less X-ray luminous on average.}
\label{fig:lxa4}
\end{figure}

Figure~\ref{fig:lxgamma}, shows \lx\ as a function of $\gamma$.  The trend reported by \citet{pel99} is confirmed here, with core galaxies having a much larger range in \lx\ than power-law galaxies which are restricted to lower values.  The same trend is seen in Figure~\ref{fig:lxa4} which shows \lx\ as a function of the deviation from from a pure ellipse, quantified by $a(4)/a$ (\citealt{ben89}), where $a$ is the semi-major axis and $a(4)$ is the fourth cosine coefficient of the Fourier transform of the azimuthal variations of the isophotes --- $a(4)$ characterises a galaxy's boxy or discy nature.  It is worth noting that the outlying point at \lx$\approx 42$ and $a(4)/a\approx 2.5$, is NGC6482, a classic fossil-group galaxy.  The high value of \lx\ is confirmed in a detailed study of Chandra data by  \citet{kho04}.  The galaxy at log\lx$<38$ in both plots is M32, the only detection in the whole sample fainter than \lx$=10^{38}$ erg s$^{-1}$.

\begin{figure}
\centering \includegraphics[scale=0.4,angle=0]{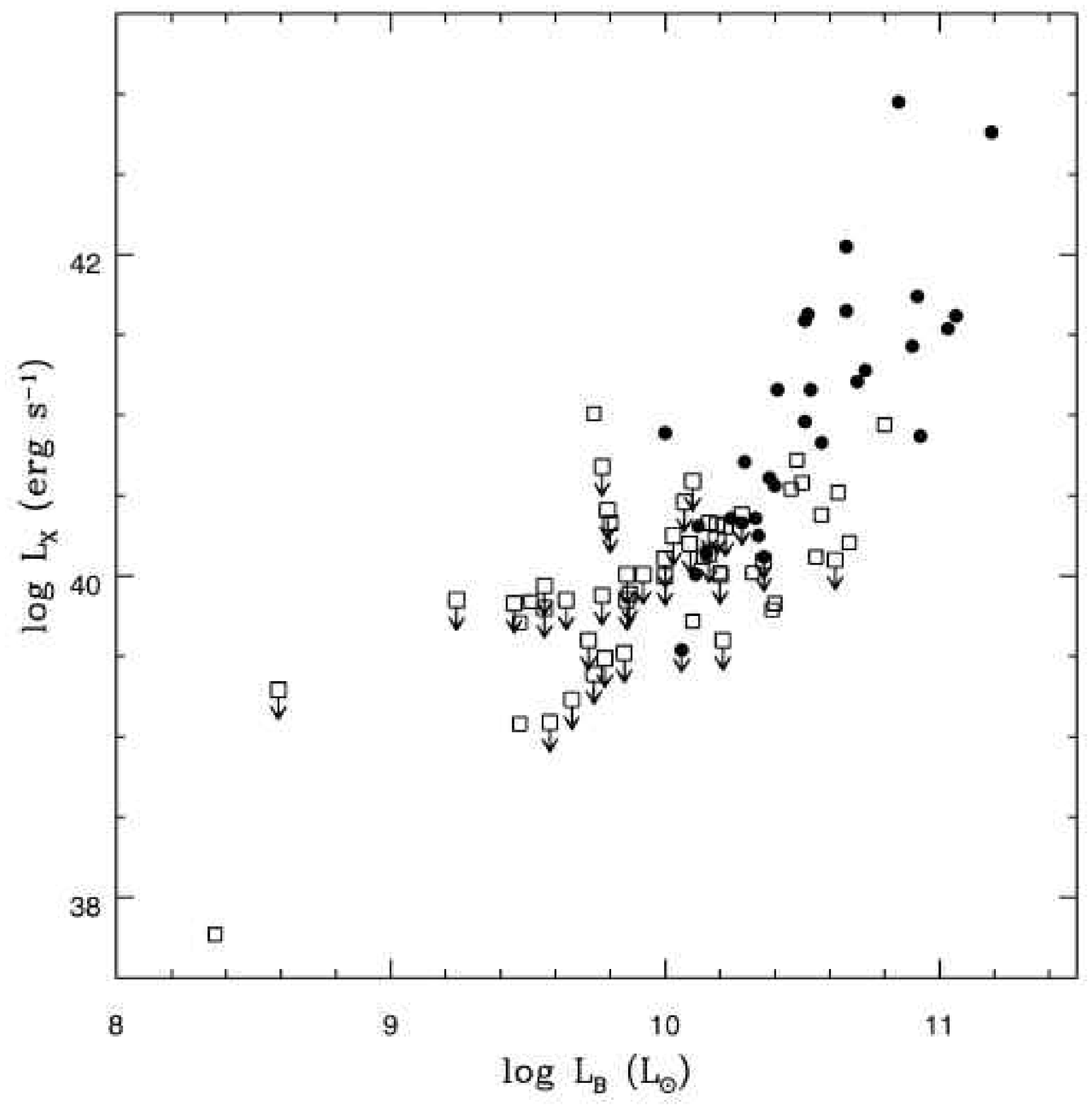}
\caption{log\lx\ as a function of log \lb, for the sample categorised according to core profile.  Symbols are as in Figure~\ref{fig:lxgamma}.  Core profile galaxies are generally optically brighter than power-law galaxies, and consequently display a steeper relation in log\lx\ vs. log\lb.}
\label{fig:lxlb_core}
\end{figure}

Categorising galaxies as core or power-law galaxies, the \lx:\lb\ relation has been replotted in Figure~\ref{fig:lxlb_core}.  The core galaxies are seen to populate the steeper part of the relation, whilst the power-law galaxies are mainly confined to the shallower relation.  This is because the power-law galaxies are generally fainter, and therefore less massive systems, whereas the core galaxies are bright and massive.   This may go someway to explaining the relation between \lx\ and $\gamma$; as both properties are related to \lb, the fact that they are correlated with each other is not surprising.  However, it seems unlikely that \lb\ alone is entirely responsible for the variation in $\gamma$ and \lx.  No power-law galaxies in the sample have \lx$>10^{41.06}$ erg s$^{-1}$, however, some of the power-law galaxies are bright enough in \lb\ that higher \lx\ may be expected (see Fig.~\ref{fig:lxlb_core}).  All the brightest power-law galaxies have lower \lx\ than would be expected from the \lx:\lb\ relation, i.e.\ there seems to be an extra factor other than \lb\ which also influences the X-ray luminosity and core-profile.  Thus it is possible that  the \lx:$\gamma$ relation may be influenced by both optical luminosity (and therefore stellar mass) and formation mechanism.

Thus the dichotomy in the shape of the low and high \lx/\lb\ galaxies may be revealing important information about the formation processes at work.  Differences in core profile shape could be due to major or minor mergers (\citealt{burkert03}; \citealt{kho03}), dissipative or dissipationless mergers (\citealt{fab97};  \citealt{ryd01}), the mass of the central black hole (\citealt{vandermarel99}), also monolithic collapse would produce ellipticals with core profiles.
It may be possible to discern the more likely of these scenarios by requiring that whatever processes are responsible for the differences in core profile, also produce the observed relations in \lx.  


\subsection{Line strength indices and galaxy age}

Absorption line strengths have been collated for the sample in order to investigate the effects of metallicity and age on the scatter of \lx/\lb.  Specifically Mg$_{2}$, H$\beta$ and H$\gamma$ line strength indices have been examined.  Mg$_{2}$ index is degenerate in age and metallicity, but has the advantage that a large sample was available.  H$\beta$ and H$\gamma$  are sensitive indicators of age, but also suffer from age-metallicity degeneracy.  The age-metallicity degeneracy considerable hampers interpretation of the results for all the line indices studied.  Values have been collated for 290 galaxies with Mg$_{2}$ (from HyperLeda), 176 galaxies with H$\beta$ and 62 galaxies with H$\gamma$ (from \citealt{trag97}; \citealt{kun00} and \citealt{den05}).

\begin{figure}
\centering \includegraphics[scale=0.4,angle=0]{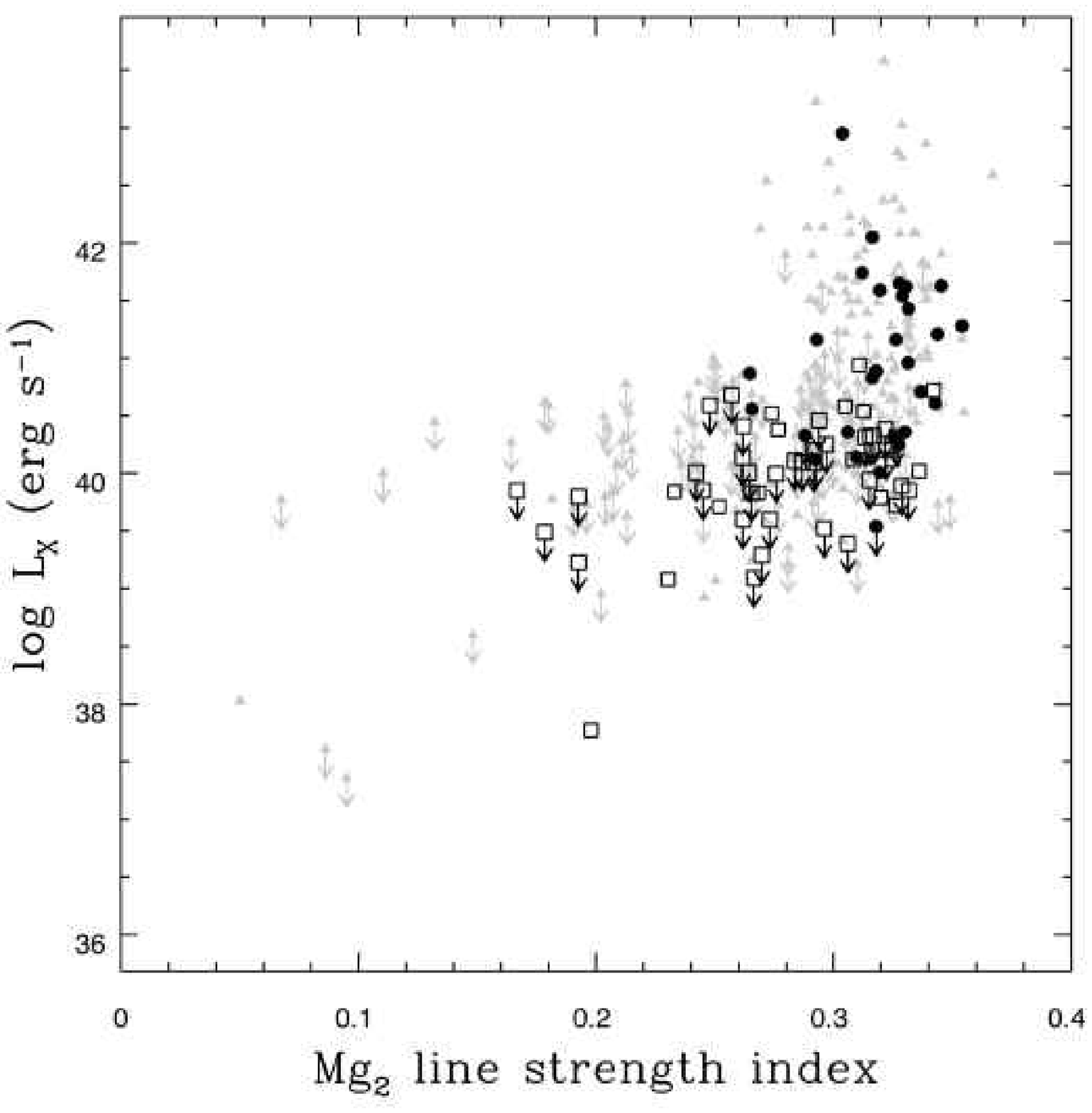}
\caption{log\lx\ as a function of Mg$_{2}$ line strength index.  Galaxies with a high Mg$_{2}$ index are generally more luminous in X-rays than galaxies with low Mg$_{2}$ index.}
\label{fig:lxmg2}
\end{figure}

A similar trend to those displayed above by core profile slope, $\gamma$, is found in the Mg$_{2}$ line strength index, and illustrated in Figure~\ref{fig:lxmg2}.  It can be seen that galaxies with weak Mg$_{2}$ absorption are generally confined to log \lx$<41$ (erg s$^{-1}$), whereas high Mg$_{2}$ galaxies having generally high X-ray emission.  Mg$_{2}$ line strength is degenerate with age and metallicity, with older stellar populations and more metal rich populations both giving rise to high Mg$_{2}$ index (\citealt{cas96}).  Thus the trend in Figure~\ref{fig:lxmg2}, could be attributed to both the age-\lx\ relation (\citealt{osul01b}) or a mass-metallicity relation, which is believed to drive the colour-magnitude relation of elliptical galaxies (\citealt{kod98}), or a combination of the two.

\begin{figure}
\centering \includegraphics[scale=0.4,angle=0]{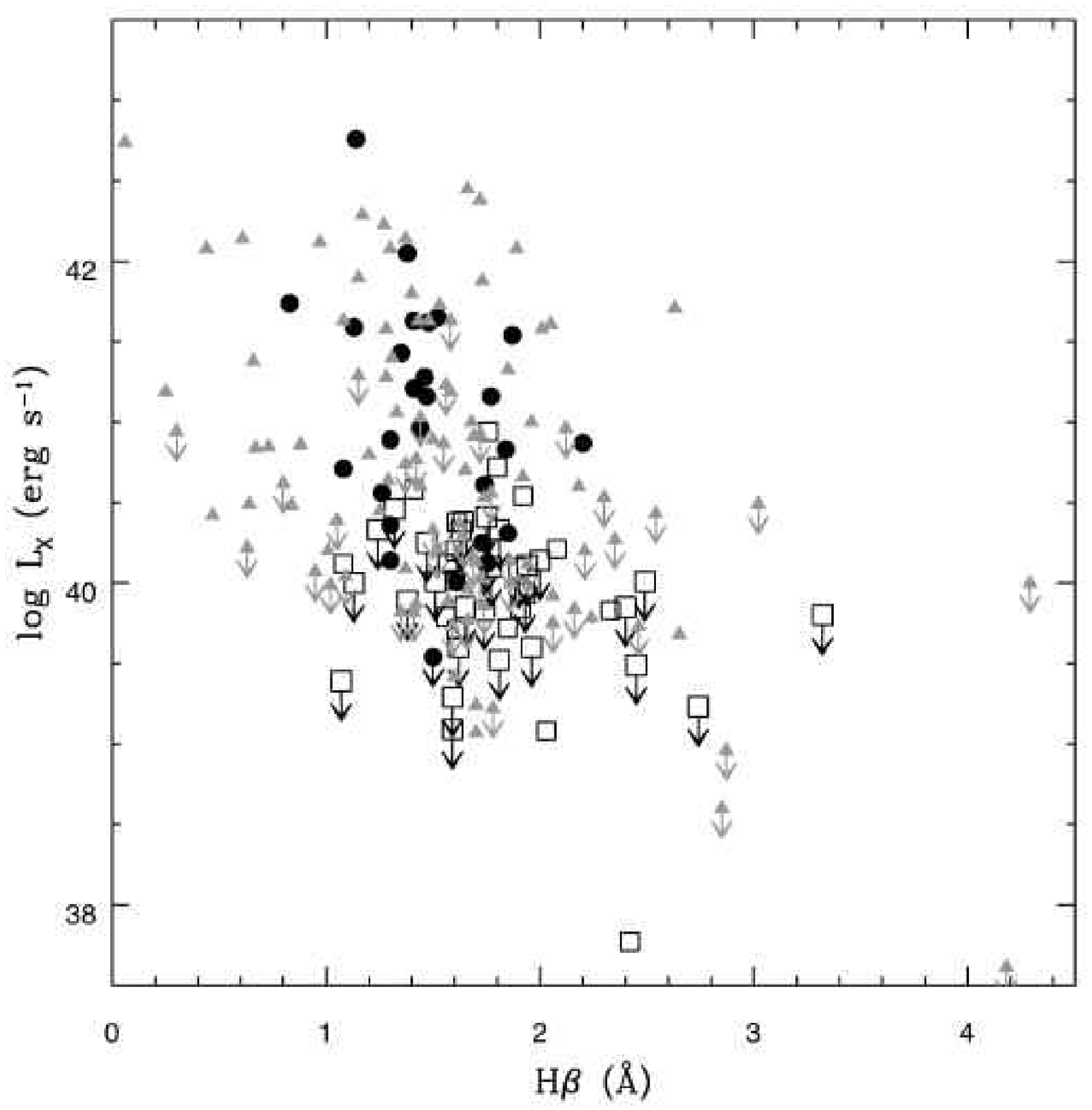}
\caption{log\lx\ as a function of H$\beta$ line strength index.   Symbols are as in Figure~\ref{fig:lxgamma}.  Galaxies with a low H$\beta$ index are generally more luminous in X-rays than galaxies with high H$\beta$ index.}
\label{fig:lxHbeta}
\end{figure}

\begin{figure}
\centering \includegraphics[scale=0.4,angle=0]{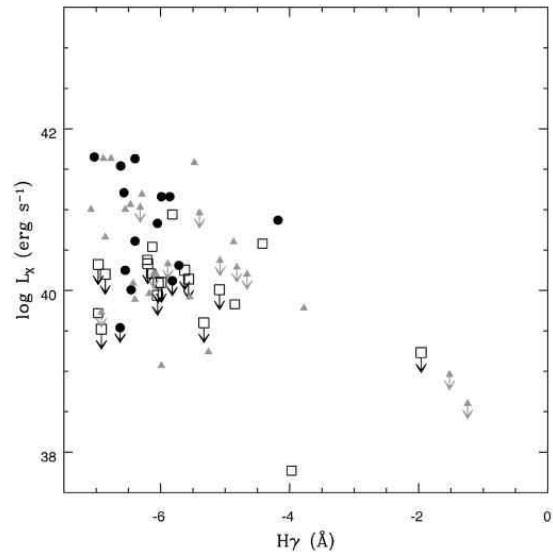}
\caption{log\lx\ as a function of H$\gamma$ line strength index.   Symbols are as in Figure~\ref{fig:lxgamma}.  Galaxies with a low H$\gamma$ index are generally more luminous in X-rays than galaxies with high H$\gamma$ index.}
\label{fig:lxHgamma}
\end{figure}

\begin{figure}
\centering \includegraphics[scale=0.4,angle=0]{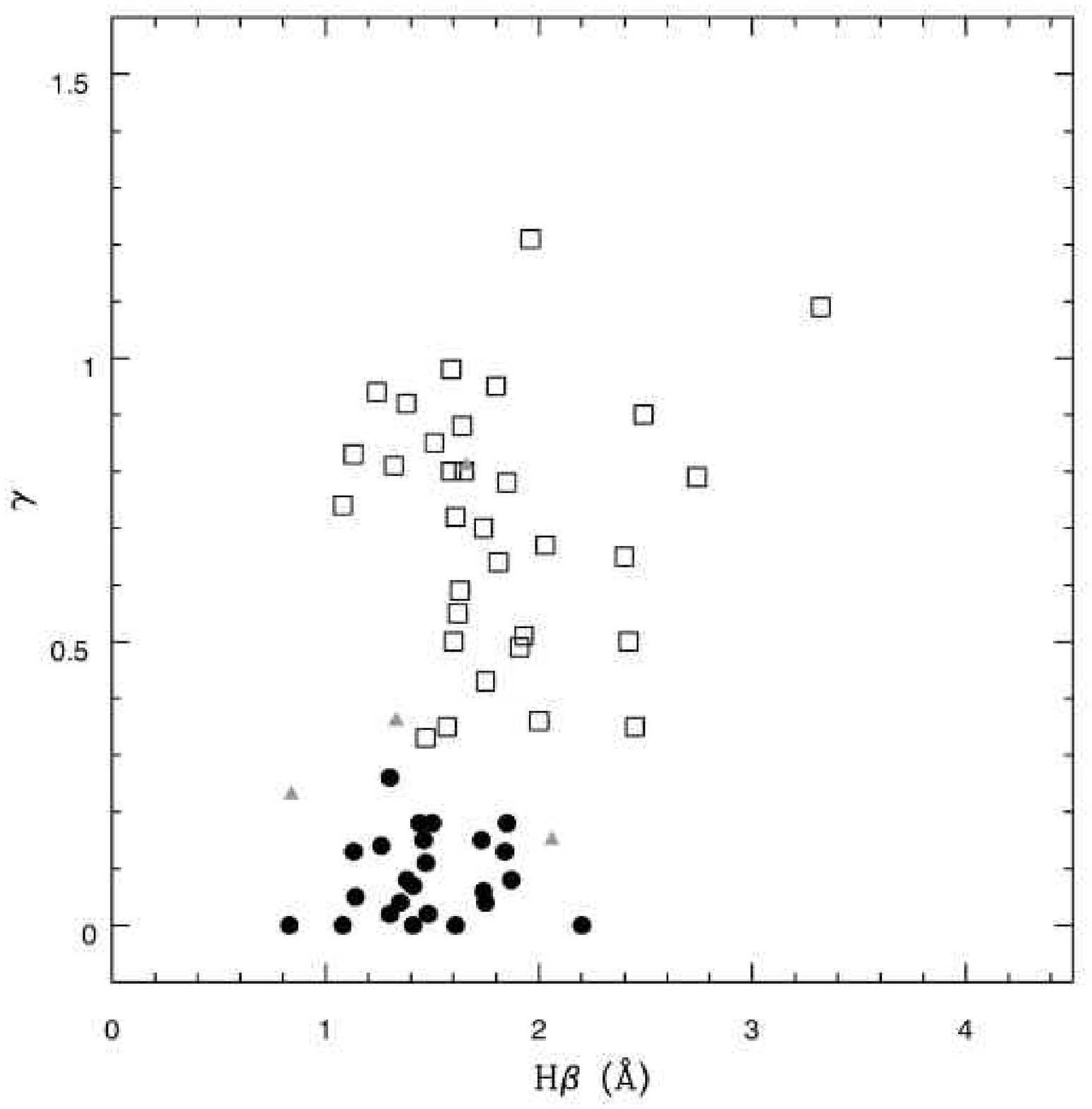}
\caption{Inner profile slope, $\gamma$,  as a function of H$\beta$ line strength index.   Symbols are as in Figure~\ref{fig:lxgamma}.  Inner profile slope is relatively insensitive to H$\beta$ index.}
\label{fig:gammaHbeta}
\end{figure}

Similarly the trends with H$\beta$ and H$\gamma$ have also been investigated.  H$\beta$ is a sensitive indicator of age (although the effects of metallicity are not negligible) but may suffer from nebular emission filling in the line in some cases, H$\gamma$ is less susceptible to this problem but is also less sensitive to age (\citealt{ter02}).  \lx\ as function of H$\beta$ and H$\gamma$ is plotted in Figures~\ref{fig:lxHbeta} and \ref{fig:lxHgamma} respectively.  Both show similar trends to that found for Mg$_{2}$,  implying that age and/or metallicity  is  important in accounting for the large \lx\ values of core galaxies.  A plot of $\gamma$ as a function of H$\beta$, reveals only a marginal trend with inner profile slope, and little difference between core and power-law galaxies as in Figure~\ref{fig:gammaHbeta}.

\begin{figure}
\centering \includegraphics[scale=0.4,angle=0]{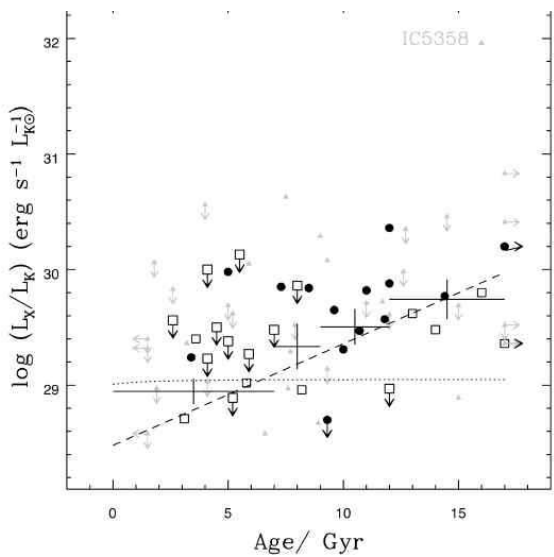}
\caption{\lx/\lb\ as a function of galaxy age.  Ages are taken from the catalogue of \protect\citet{ter02}.   Symbols are as in Figure~\ref{fig:lxgamma}. The dashed line is a fit to the data, while the dotted line shows the expected change in \lx/\lk\ with age due to decreasing \lk\ alone. Large crosses show the mean \lx/\lk\ in the four age bins described in the text.}
\label{fig:age}
\end{figure}

By using a combination of spectral lines it is possible to try and break
the age-metallicity degeneracy.  For example, \citet{ter02} present a
catalogue of galaxy ages using the H$\beta$ line index and the combination
index [MgFe].  \citet{osul01b} have used this catalogue to investigate the
correlation of \lx/\lb\ and find an increase in \lx\ with galaxy age.
Figure~\ref{fig:age} shows the equivalent plot in terms of \lk. The trend
is very similar to that found for \lx/\lb, with \lx/\lk\ increasing in
older galaxies. The correlation has a significance of at least 99.7\%, and
line fits to the data have slopes of 0.088 $\pm$ 0.021 (EM) and 0.107 $\pm$
0.025 (BJ). The former is shown as the dashed line in Figure
~\ref{fig:age}. Excluding IC~5358, which has an unusually high \lx/\lk,
flattens the fits slightly, giving slopes of 0.0704 $\pm$ 0.0185 (EM) and
0.0989 $\pm$ 0.0228 (BJ). As a further test of the nature of the
correlation, we binned the sample into subsets by age, insisting that there
be a minimum of 7 X-ray detected galaxies in each bin. The mean \lx/\lk\ 
(calculated using the Kaplan-Meier estimator) for these bins are shown as
large crosses in Figure~\ref{fig:age}. A part of the trend in \lx/\lk\ with
age will arise from changes in the optical properties of the stellar
population over time. We have estimated the change in \lk\ with age from
the models of \citet{bru03}, assuming a galaxy with solar metallicity in
which 10\% of the mass of stars is formed in a starburst, and 90\% are old
with negligible change of \lk\ with increasing age. The dotted curve shows
the expected change in \lx/\lk\ for such a galaxy, given an unchanging \lx.
As expected, the contribution to the trend in \lx/\lk\ from evolution of
the stellar population is negligible. \citet{osul01b} used a slightly
larger number of galaxies than is included in this sample (77 compared to
67), and found better agreement between the two line fitting techniques,
but the results for \lx/\lk\ are clearly very similar to those found for
\lx/\lb, in that the mean \lx/\lk\ increases steadily over the whole range
of age in the sample.

\subsection{Scatter as a function of galactic nuclear activity}

\begin{figure}
\centering \includegraphics[scale=0.4,angle=0]{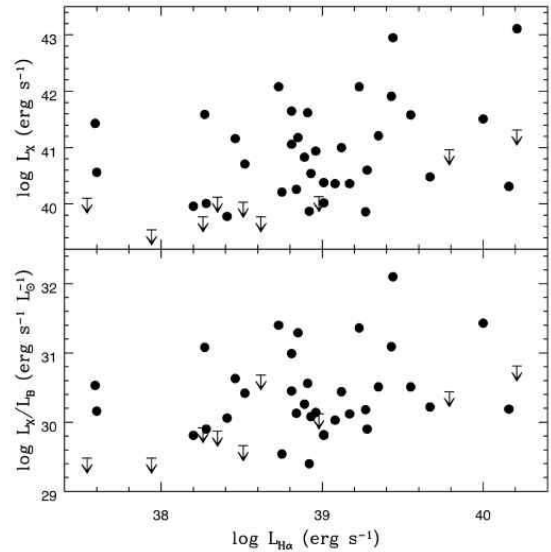}
\caption{log\lx\ as a function of log$L_{{\rm H}\alpha}$ (top) and log\lx/\lb\ as a function of log$L_{{\rm H}\alpha}$ (bottom).  log\lx\ and log\lx/\lb\ are insensitive to the amount of nuclear H$\alpha$ emission present.  The H$\alpha$ emission has had the starlight and continuum contribution subtracted.}
\label{fig:lxHalpha}
\end{figure}

The presence of an active galactic nucleus (AGN) may affect the X-ray  (e.g.\ \citealt{ho01}) and optical emission of a galaxy, and thus a sample which contains galaxies both possessing and lacking AGN may increase the scatter of the \lx:\lb\ relation.  All known AGN have been excluded from the sample, but it is possible that some weaker AGN remain.  In order to quantify the amount of galactic nuclear activity the H$\alpha$ emission line strength, indicative of a Seyfert I galaxy, has been used.  \citet{ho97a} present H$\alpha$ luminosities for the nuclei of 418 galaxies from with the starlight contribution has been subtracted, hence  any residual H$\alpha$ emission will be from nuclear activity or star-formation.  Of the galaxies presented by \citet{ho97a}, 45 overlap with the sample presented here.  This sample has been matched to the galaxies of \citet{osul01}, and the dependence of log \lx\ and log \lx$/$\lb\ on log $L_{{\rm H}\alpha}$ is shown in Figure~\ref{fig:lxHalpha}.  Although the sample is only small it can be seen that there is at best only a weak trend of \lx\ and \lx/\lb\ with galactic nuclear activity.  The range in \lx\ and \lx/\lb\ is similar to the range of the entire sample.  Thus it appears that galactic nuclear activity is not responsible for the large scatter observed in the \lx:\lb\ relation.

\section{Discussion}
\label{sec:discussion}

The relationship between X-ray luminosity and NIR luminosity for early-type galaxies has been examined.  The results are consistent with
previous work studying the relationship between X-ray luminosity and optical luminosity (\citealt{osul01}).   The \lx:\lb, \lx:\lh\ and
\lx:\lk\ relations bear the same
overall trend, becoming steeper for galaxies that are bright in the optical or NIR.  This is consistent with previous interpretations in which more
massive galaxies contain larger amounts of hot gas (see \citealt{mat03} for a review).

The scatter of the relations is the same in the \emph{B}, \emph{H} and \emph{Ks} bands.  This is probably largely because early-type galaxies generally contain very
little dust and are composed of old, quiescent stellar populations.  Any scatter introduced  by residual star-formation or dust present in
early-types,  must be insignificant compared to the large scatter of the X-ray luminosities.

The scatter and slope of the relations has been investigated as a function of environment.  There is no clear trend with environment, either in terms of
cluster, group or field membership or in terms of local galaxy density.  
Brightest group galaxies display a steeper relation and significantly higher \lx/\lb, \lx/\lh\ and \lx/\lk, than other types of galaxies, perhaps indicative of a different formation mechanism or their location in the densest regions of the IGM, in agreement with \citet{osul01} and \citet{hel01}.  There may also be a slight increase in the \lx/\lb\ etc.\ values for group galaxies in general again suggestive that the IGM is influential on the X-ray emission from galaxies through processes such as accretion or stifling of galactic winds.  However cluster galaxies, which would be subject to similar processes are consistent with \lx/\lb\ values of field galaxies, making any firm conclusions difficult to draw.  It is possible that differences between group and cluster galaxies, such as merger history, ram-pressure stripping, etc.\ may be responsible for the lower \lx/\lb\ of cluster galaxies.  For example, it is known that group galaxies are involved in  significantly more galaxy-galaxy interactions than cluster galaxies, since the higher velocity dispersion of clusters make the chances of galaxy-galaxy interaction much smaller (\citealt{zep93}).

Galaxies with a shallow inner core profile cover a wider range in \lx\ than power-law galaxies with a steep inner profile, which are generally restricted to log\lx$<41$ (erg s$^{-1}$).  This is in agreement with previous work by \citet{pel99}.  Furthermore the core profile galaxies are observed mainly to occupy the steeper part of the \lx:\lb\ relation, while the power-law galaxies are generally on the shallower part of the relation.  This is suggestive that whatever is causing the break in the \lx:\lb\ relation is also responsible for the observed differences between core and power-law galaxies, such as rotation velocity, luminosity etc.    However, this it is difficult to test this as the large scatter in both relations makes a comparison of the \emph{B} band luminosities of the breaks hard.  An obvious difference is the mass of the systems, but different formation mechanisms are also often speculated as the cause of the difference between power-law and core galaxies (e.g.\ \citealt{fab97}; \citealt{ryd01}; \citealt{burkert03}; \citealt{kho03}).

The effects of age and metallicity were investigated via absorption line strengths.  Mg$_{2}$ index, which is degenerate in age and metallicity, shows a clear demarcation with high index galaxies having a large spread in \lx\ while lower index galaxies are confined to lower values of \lx.  Similar trends are also seen for H$\beta$ and H$\gamma$ index which are less degenerate, and more sensitive to age (\citealt{ter02}).  Although this provides support for a relation between age and \lx, as reported by \citealt{osul01b}, it seems likely that the large spread in \lx\ for high Mg$_{2}$ index galaxies is also partly due to a higher mass and subsequent higher metallicity.  The age-metallicty degeneracy can be broken using combinations of emission lines.  Using the spectroscopic ages of \citet{ter02} reduces the size of the sample, but reveals a strong correlation between \lx/\lk\ and age.  The contribution to this correlation due to the dimming of \lk\ over time is negligible, supporting the results and conclusions of \citet{osul01b}.

Known AGN were removed from the sample before any statistical analysis was performed.  However it is possible that some galaxies with a low level of nuclear activity remain in the sample.  Galactic nuclear activity has been quantified using H$\alpha$ emission line strength, as reported by \citet{ho97a} after subtraction of the contribution from starlight.  Broad H$\alpha$ emission lines are a classic indication of a Seyfert or LINER (low-ionisation nuclear emission-line region) galaxy .  Seyfert and LINER galaxies are generally spirals, but occasionally are found in ellipticals.  The lack of correlation of either \lx\ or log\lx/\lb\ with H$\alpha$ emission is suggestive that galactic nuclear activity is not very influential on the observed scatter of the \lx:\lb\ relation.  This is corroborated by the similar range in values of \lx\ and log\lx/\lb\ displayed by the H$\alpha$ emitting galaxies and the whole sample.

In conclusion, the origin of the large scatter in \lx/\lb\ remains elusive.  It is clear that mass, age, metallicity and core-profile are all linked with the range in \lx\ displayed by galaxies, with more massive and older galaxies having a wider range in \lx\ and cuspy profiles.   The underlying cause of these differences may well be linked with formation and merger history as well as the  environment of the galaxies, but disentangling the effects is very problematic. Conversely it appears that star-formation and recent post-merger objects do not significantly contribute to the scatter, and thus studies of the optical-X-ray luminosity relations are a valid choice.   Similarly the local density and environment of early-type galaxies do not appear to affect the scatter.

\section*{Acknowledgments}

We extend our warmest thanks to Duncan Forbes for his helpful comments on an earlier draft of this paper, and to Trevor Ponman for his suggestions on this work.  Many thanks, also, to Doug Burke for discussions and software help.
We thank the referee for useful comments which have improved this paper.

This publication makes use of data products from the Two Micron All Sky Survey, which is a joint project of the
University of Massachusetts and the Infrared Processing and Analysis Center/California Institute of Technology, funded
by the National Aeronautics and Space Administration and the National Science Foundation. 

This research has made use of the NASA/IPAC Extragalactic Database (NED) which is operated by the Jet Propulsion
Laboratory, California Institute of Technology, under contract with the National Aeronautics and Space Administration.

This research has also made use of the HyperLeda archives available at http://leda.univ-lyon1.fr/.

SCE acknowledges PPARC support whilst researching this work.  EO'S acknowledges support from NASA grant NNG04GF19G and Chandra Award Number AR4-5102X.

\bibliographystyle{mn2e}
\bibliography{clusters}

\appendix

\section{The catalogue}

We provide a catalogue of the data used in this paper for the convenience of the reader.  The catalogue is also available online at http://hea-www.harvard.edu/$\sim$ejos/catalogue.html.  We emphasise that none of the data presented in this table are new, but are compilations of data from the literature; references for each source are given as footnotes in the table.  Where data in the main part of the paper were from a single source (e.g.\ age) we have excluded these from the table.

\onecolumn
\begin{longtable}{lcccccccccccc}

\caption{Summary of the data and sources used in the paper.}
\label{tab:bigtab}

\\ \hline 
Name & Distance$^{a}$ & log\lb$^{b}$ & log\lk$^{c}$ & log\lh$^{c}$ & log\lx$^{d}$ & Env.1$^{e}$ & Env.2$^{f}$ & Mg$_{2}$$^{g}$ & 100.a4/a$^{h}$ & $\gamma^{i}$ & H$\beta^{j}$ & H$\gamma^{j}$\\
& Mpc & \lbsun & \lksun & \lhsun & erg s$^{-1}$ & & & & & & & \\ \hline
\endfirsthead

\multicolumn{13}{l}{\emph{continued from previous page}}\\
\\ \hline 
Name & Distance$^{a}$ & log\lb$^{b}$ & log\lk$^{c}$ & log\lh$^{c}$ & log\lx$^{d}$ & Env.1$^{e}$ & Env.2$^{f}$ & Mg$_{2}$$^{g}$ & 100.a4/a$^{h}$ & $\gamma^{i}$ & H$\beta^{j}$ & H$\gamma^{j}$\\ 
& Mpc & \lbsun & \lksun & \lhsun & erg s$^{-1}$ & & & & & & & \\ \hline
\endhead

\\ \hline 
\endfoot

\\ \hline 
\multicolumn{13}{l}{The sources for each column are listed in order of preference.}\\
\multicolumn{13}{l}{$^{a}$ \protect\citet{pru96}, HyperLeda.} \\
\multicolumn{13}{l}{$^{b}$ $B_{{\rm T}}$ from LEDA, $B_{{\rm T}}$ from NED, $m_{{\rm B}}$ from LEDA, $m_{{\rm B}}$ from NED.  Values derived from $m_{{\rm B}}$ are marked with an asterisk.} \\
\multicolumn{13}{l}{$^{c}$ \protect\citet{jar03}, \protect\citet{jar00}.  Apparent magnitudes were converted to luminosities}\\
\multicolumn{13}{l}{ using the distances in column 2 (see the text for details).} \\
\multicolumn{13}{l}{$^{d}$ All values are from \protect\citet{osul01}, however, some of these were derived from the catalogues \protect\citet{beu99},}\\\multicolumn{13}{l}{\protect\citet{fab92} and \protect\citet{rob91}.} \\
\multicolumn{13}{l}{$^{e}$ Environment as given in \protect\citet{osul01}, F=field galaxy, G=group member (based on \protect\citealt{gar93}),}\\
\multicolumn{13}{l}{C=cluster member (based on \protect\citealt{abe89} and \protect\citealt{faber89}),}\\
\multicolumn{13}{l}{D=distance$>70$Mpc (for which environment cannot be accurately determined).}\\
\multicolumn{13}{l}{$^{f}$ Brightest group galaxies (G, from \protect\citealt{gar93}), brightest cluster galaxies (C, from \protect\citealt{abe89} and \protect\citealt{faber89}),}\\
\multicolumn{13}{l}{dwarfs (D, \lb$<10^{9}$\lbsun) and AGN (A, from \protect\citealt{ver01}) as given in  \protect\citet{osul01}.}\\
\multicolumn{13}{l}{$^{g}$ HyperLeda, where more than one value was available an average value is given.}\\
\multicolumn{13}{l}{$^{h}$ \protect\citet{pel99}, \protect\citet{fab97}, \protect\citet{ben89}.}\\
\multicolumn{13}{l}{$^{i}$ \protect\citet{pel99}, \protect\citet{rest01}, \protect\citet{rav02}, \protect\citet{fab97}.}\\
\multicolumn{13}{l}{$^{j}$ \protect\citet{den05}, \protect\citet{kun00}, \protect\citet{trag97}.}\\
\hline
\endlastfoot

ESO101-14 & 30.12 & 9.93$^{*}$ & 10.84 & 10.74 & $\le$41.02 & F & - & - & - & - & - & -\\
ESO107-4 & 38.89 & 10.22 & 10.88 & 10.77 & $\le$40.94 & F & - & 0.249 & - & - & - & -\\
ESO137-6 & 69.75 & 10.56 & 11.74 & 11.62 & 42.08 & G & G & 0.307 & - & - & - & -\\
ESO137-8 & 47.95 & 10.42$^{*}$ & 11.42 & 11.32 & 41.22 & G & - & 0.305 & - & - & - & -\\
ESO137-10 & 42.27 & 10.46$^{*}$ & 11.21 & 11.12 & 40.94 & G & G & 0.292 & - & - & - & -\\
ESO138-5 & 35.39 & 10.16$^{*}$ & 11.11 & 11.03 & $\le$41.18 & G & - & - & - & - & - & -\\
ESO148-17 & 38.36 & 10.04 & 10.72 & 10.63 & $\le$40.68 & F & - & - & - & - & - & -\\
ESO183-30 & 33.59 & 10.18 & 11.07 & 10.99 & $\le$41.00 & G & - & 0.250 & - & - & - & -\\
ESO185-54 & 56.36 & 10.84$^{*}$ & 11.68 & 11.58 & 41.36 & G & G & 0.332 & - & - & - & -\\
ESO208-21 & 10.36 & 9.34 & 10.23 & 10.13 & 39.73 & F & - & 0.322 & - & - & - & -\\
ESO243-45 & 100.91 & 10.84$^{*}$ & 11.47 & 11.39 & $\le$41.91 & D & - & - & - & - & - & -\\
ESO273-2 & 3.20 & 7.54 & 8.37 & 8.26 & $\le$38.64 & F & D & - & - & - & - & -\\
ESO286-50 & 33.31 & 9.76 & 10.34 & 10.24 & $\le$40.53 & F & - & - & - & - & - & -\\
ESO306-17 & 139.95 & 11.15$^{*}$ & 11.83 & 11.69 & 43.33 & D & - & - & - & - & - & -\\
ESO381-29 & 32.65 & 9.78 & 10.48 & 10.39 & $\le$40.59 & F & - & 0.180 & - & - & - & -\\
ESO400-30 & 30.17 & 9.76 & 9.80 & 9.70 & $\le$40.45 & F & - & 0.132 & - & - & - & -\\
ESO428-11 & 10.49 & 9.07 & 9.93 & 9.83 & $\le$39.63 & F & - & - & - & - & - & -\\
ESO443-24 & 65.97 & 10.67 & 11.58 & 11.48 & 41.50 & G & G & 0.307 & - & - & - & -\\
ESO495-21 & 9.16 & 9.13$^{*}$ & 9.67 & 9.57 & 39.56 & F & - & - & - & - & - & -\\
ESO507-21 & 40.23 & 10.51$^{*}$ & 10.84 & 10.73 & 40.93 & G & - & 0.326 & - & - & - & -\\
ESO552-20 & 123.49 & 11.04$^{*}$ & 11.84 & 11.71 & 42.59 & D & - & 0.367 & - & - & - & -\\
ESO553-2 & 61.88 & 10.42 & 11.16 & 11.07 & 41.50 & F & - & 0.310 & - & - & - & -\\
ESO565-30 & 132.99 & 11.05$^{*}$ & 11.71 & 11.61 & 42.37 & D & - & 0.321 & - & - & - & -\\
IC310 & 63.39 & 10.54 & 11.29 & 11.17 & 42.54 & G & A & 0.272 & - & - & - & -\\
IC989 & 101.33 & 10.60 & 11.52 & 11.30 & $\le$41.55 & D & C & - & - & - & - & -\\
IC1024 & 21.68 & 9.31$^{*}$ & 9.99 & 9.86 & $\le$40.20 & G & - & - & - & - & - & -\\
IC1459 & 18.88 & 10.37 & 11.18 & 11.08 & 40.71 & G & G & 0.335 & - & - & - & -\\
IC1531 & 100.69 & 10.87$^{*}$ & 11.53 & 11.40 & 41.60 & D & - & - & - & - & - & -\\
IC1625 & 86.20 & 10.90 & 11.50 & 11.41 & 41.75 & D & C & 0.313 & - & - & - & -\\
IC1633 & 93.81 & 11.09 & 11.93 & 11.83 & 42.79 & D & - & 0.327 & - & - & - & -\\
IC1729 & 18.09 & 9.29 & 10.00 & 9.92 & $\le$40.00 & F & - & - & - & - & - & -\\
IC1860 & 90.15 & 10.62 & 11.56 & 11.42 & 42.71 & D & - & - & - & - & - & -\\
IC2006 & 18.11 & 9.88 & 10.47 & 10.37 & $\le$41.03 & C & - & 0.296 & - & - & 1.44 & -6.32\\
IC2035 & 16.52 & 9.64 & 10.24 & 10.13 & $\le$40.62 & F & - & 0.179 & - & - & - & -\\
IC2311 & 22.11 & 9.88 & 10.69 & 10.60 & $\le$40.22 & F & - & 0.262 & - & - & - & -\\
IC2533 & 31.45 & 10.00 & 10.85 & 10.77 & $\le$40.44 & F & - & - & - & - & - & -\\
IC2552 & 38.37 & 10.00$^{*}$ & 10.93 & 10.81 & $\le$40.69 & G & - & - & - & - & - & -\\
IC2597 & 58.34 & 10.58 & 11.47 & 11.37 & $\le$41.03 & C & - & 0.318 & - & - & - & -\\
IC3896 & 25.29 & 9.97$^{*}$ & 11.04 & 10.95 & $\le$40.50 & F & - & 0.320 & - & - & - & -\\
IC3986 & 59.49 & 10.41$^{*}$ & 11.36 & 11.26 & 40.30 & G & - & 0.309 & - & - & - & -\\
IC4197 & 38.63 & 9.95 & 10.81 & 10.71 & $\le$40.73 & G & - & 0.28 & - & - & - & -\\
IC4296 & 47.56 & 10.90 & 11.70 & 11.56 & 41.53 & G & G & 0.339 & - & - & - & -\\
IC4329 & 58.83 & 10.86$^{*}$ & 11.67 & 11.55 & $\le$41.82 & G & G & 0.338 & 0.80 & - & - & -\\
IC4765 & 58.20 & 10.79$^{*}$ & 11.61 & 11.48 & 41.83 & G & G & 0.338 & - & - & - & -\\
IC4797 & 33.31 & 10.31 & 11.17 & 11.07 & $\le$40.99 & G & G & 0.302 & - & - & - & -\\
IC4889 & 29.51 & 10.42 & 11.04 & 10.96 & $\le$40.80 & F & - & 0.258 & 1.00 & - & - & -\\
IC4943 & 34.67 & 9.90 & 10.50 & 10.45 & $\le$40.64 & G & - & 0.251 & - & - & - & -\\
IC5181 & 24.63 & 9.97 & 10.86 & 10.77 & $\le$40.28 & G & - & - & - & - & - & -\\
IC5269 & 24.52 & 9.69$^{*}$ & 10.35 & 10.26 & $\le$40.46 & G & - & - & - & - & - & -\\
IC5358 & 113.09 & 10.86 & 11.62 & 11.54 & 43.58 & D & - & 0.321 & - & - & - & -\\
NGC57 & 55.21 & 10.61 & 11.36 & 11.25 & 41.65 & F & - & - & - & - & - & -\\
NGC127 & 48.53 & 9.43 & 10.26 & 10.13 & $\le$41.16 & G & - & - & - & - & - & -\\
NGC130 & 48.53 & 9.60 & 10.17 & 10.08 & $\le$41.18 & G & - & - & - & - & - & -\\
NGC147 & 0.65 & 7.92 & 8.09 & 8.04 & $\le$37.45 & G & D & - & - & - & - & -\\
NGC185 & 0.62 & 8.07 & 8.31 & 8.26 & $\le$37.36 & G & D & 0.0950 & - & - & - & -\\
NGC205 & 0.72 & 8.40 & 8.83 & 8.72 & $\le$37.61 & G & D & 0.0860 & - & - & 4.18 & -\\
NGC221 & 0.72 & 8.36 & 9.02 & 8.92 & 37.77 & G & D & 0.198 & 0.00 & 0.50 & 2.42 & -3.97\\
NGC227 & 71.01 & 10.65 & 11.41 & 11.30 & $\le$41.23 & D & - & 0.302 & - & - & 1.56 & -\\
NGC315 & 58.88 & 11.07 & 11.71 & 11.61 & 41.58 & G & G & 0.299 & -0.31 & - & 2.01 & -5.48\\
NGC383 & 56.49 & 10.86 & 11.46 & 11.35 & 41.38 & G & - & 0.307 & - & - & 0.66 & -\\
NGC410 & 56.75 & 10.82 & 11.50 & 11.40 & 41.91 & G & - & 0.345 & - & - & - & -\\
NGC439 & 74.60 & 10.94$^{*}$ & 11.62 & 11.50 & 41.71 & D & - & 0.302 & - & - & - & -\\
NGC499 & 55.21 & 10.57 & 11.34 & 11.23 & 42.29 & G & G & 0.329 & - & - & 1.17 & -\\
NGC507 & 67.19 & 10.96 & 11.68 & 11.53 & 42.70 & G & G & 0.298 & - & - & - & -\\
NGC529 & 65.96 & 10.57 & 11.34 & 11.27 & 40.60 & G & - & 0.298 & - & - & 1.44 & -\\
NGC533 & 63.68 & 10.90 & 11.58 & 11.47 & 42.23 & G & G & 0.307 & 0.00 & - & 1.27 & -\\
NGC541 & 63.39 & 10.66 & 11.23 & 11.13 & 40.84 & G & - & 0.316 & - & - & 0.67 & -\\
NGC547 & 63.39 & 10.92 & 11.56 & 11.45 & 40.70 & G & - & 0.321 & - & - & - & -\\
NGC568 & 73.24 & 10.49 & 11.25 & 11.16 & 41.49 & D & - & 0.293 & - & - & - & -\\
NGC584 & 22.18 & 10.36 & 11.12 & 11.05 & $\le$40.09 & G & G & 0.292 & 1.50 & - & 1.90 & -6.04\\
NGC596 & 22.28 & 10.21 & 10.85 & 10.77 & $\le$39.60 & G & - & 0.262 & 1.30 & 0.55 & 1.62 & -\\
NGC636 & 22.28 & 10.00 & 10.67 & 10.58 & $\le$40.11 & G & - & 0.284 & 0.80 & - & 1.94 & -\\
NGC708 & 55.21 & 10.74 & 11.40 & 11.36 & 43.03 & C & C & 0.329 & - & - & - & -\\
NGC720 & 20.80 & 10.38 & 11.08 & 10.98 & 40.61 & G & G & 0.343 & 0.35 & 0.06 & 1.74 & -6.40\\
NGC741 & 61.09 & 10.90 & 11.60 & 11.51 & 41.73 & G & G & 0.326 & - & - & 1.53 & -\\
NGC777 & 55.21 & 10.68 & 11.49 & 11.37 & 42.08 & G & G & 0.335 & -0.30 & - & 1.89 & -\\
NGC821 & 20.99 & 10.16 & 10.83 & 10.74 & $\le$40.33 & F & - & 0.315 & 2.50 & 0.64 & 1.81 & -6.20\\
NGC855 & 9.33 & 8.89 & 9.35 & 9.26 & $\le$39.77 & F & D & 0.0675 & - & - & - & -\\
NGC1016 & 73.79 & 10.95 & 11.65 & 11.55 & 41.28 & D & C & 0.324 & - & - & 1.28 & -\\
NGC1052 & 17.70 & 10.12 & 10.86 & 10.77 & 40.31 & G & - & 0.325 & 0.00 & 0.18 & 1.85 & -5.72\\
NGC1167 & 67.67 & 10.50$^{*}$ & 11.55 & 11.48 & $\le$41.31 & G & - & - & - & - & - & -\\
NGC1172 & 28.71 & 10.10 & 10.58 & 10.51 & $\le$40.59 & F & - & 0.248 & - & 1.01 & - & -\\
NGC1199 & 28.71 & 10.24 & 10.84 & 10.73 & 39.42 & G & G & 0.304 & - & - & 1.60 & -\\
NGC1201 & 20.67 & 10.16$^{*}$ & 10.91 & 10.83 & $\le$40.26 & G & - & 0.305 & - & - & - & -\\
NGC1209 & 28.71 & 10.19 & 10.93 & 10.85 & $\le$40.62 & G & - & 0.296 & - & - & 0.80 & -\\
NGC1316 & 18.11 & 10.93 & 11.63 & 11.52 & 40.87 & C & - & 0.265 & 1.00 & 0.00 & 2.20 & -4.18\\
NGC1332 & 19.68 & 10.27 & 11.12 & 11.02 & 40.53 & G & - & 0.355 & - & - & 1.74 & -\\
NGC1336 & 18.11 & 9.46 & 9.94 & 9.82 & $\le$40.29 & C & - & 0.211 & - & - & 1.64 & -4.82\\
NGC1339 & 18.11 & 9.73 & 10.39 & 10.29 & $\le$40.21 & C & - & 0.308 & - & - & 1.52 & -6.11\\
NGC1344 & 18.11 & 10.30 & 10.90 & 10.82 & $\le$39.48 & C & - & - & - & - & - & -\\
NGC1351 & 18.11 & 9.78 & 10.35 & 10.28 & $\le$40.33 & C & - & 0.284 & - & - & 1.50 & -5.89\\
NGC1366 & 18.11 & 9.56 & 10.26 & 10.18 & $\le$40.32 & C & - & 0.267 & - & - & - & -\\
NGC1374 & 18.11 & 9.98 & 10.60 & 10.52 & 39.89 & C & - & 0.319 & - & - & 1.57 & -6.40\\
NGC1375 & 18.11 & 9.58 & 10.02 & 9.91 & $\le$38.60 & C & - & 0.148 & - & - & 2.85 & -1.24\\
NGC1379 & 18.11 & 10.09 & 10.57 & 10.49 & 39.24 & C & - & 0.265 & - & - & 1.70 & -5.26\\
NGC1380 & 18.11 & 10.46 & 11.12 & 11.02 & 40.09 & C & - & 0.301 & - & - & 1.37 & -6.43\\
NGC1380A & 18.11 & 9.65 & 10.03 & 9.93 & $\le$38.96 & C & - & 0.202 & - & - & 2.87 & -1.52\\
NGC1381 & 18.11 & 9.79 & 10.49 & 10.41 & 39.07 & C & - & 0.250 & - & - & 1.70 & -5.99\\
NGC1387 & 18.11 & 10.03 & 10.89 & 10.76 & 40.48 & C & - & - & - & - & - & -\\
NGC1389 & 18.11 & 9.71$^{*}$ & 10.41 & 10.32 & $\le$40.08 & C & - & 0.236 & - & - & - & -\\
NGC1399 & 18.11 & 10.52 & 11.34 & 11.24 & 41.63 & C & C & 0.345 & 0.10 & 0.07 & 1.41 & -6.40\\
NGC1395 & 20.51 & 10.44 & 11.21 & 11.11 & 40.89 & G & G & 0.325 & - & - & 1.50 & -\\
NGC1400 & 20.51 & 10.14 & 10.85 & 10.76 & 40.12 & G & G & 0.325 & 0.00 & 0.35 & 1.57 & -\\
NGC1404 & 18.11 & 10.35 & 11.14 & 11.03 & 41.19 & C & - & 0.330 & - & - & 1.58 & -6.29\\
NGC1407 & 20.61 & 10.58 & 11.30 & 11.18 & 41.00 & G & G & 0.340 & -0.30 & - & 1.96 & -6.55\\
NGC1411 & 10.56 & 9.34 & 10.10 & 10.06 & $\le$39.63 & G & - & 0.213 & - & - & - & -\\
NGC1419 & 18.11 & 9.34 & 9.91 & 9.84 & $\le$40.37 & C & - & 0.235 & - & - & 1.62 & -5.08\\
NGC1426 & 20.61 & 9.92 & 10.51 & 10.42 & $\le$40.01 & G & - & 0.264 & 0.00 & 0.85 & 1.51 & -\\
NGC1439 & 20.61 & 10.00 & 10.55 & 10.46 & $\le$40.00 & G & - & 0.276 & -0.10 & 0.83 & 1.13 & -\\
NGC1497 & 84.12 & 10.41$^{*}$ & 11.41 & 11.27 & $\le$41.33 & D & C & - & - & - & - & -\\
NGC1510 & 10.01 & 8.80 & 9.21 & 9.29 & $\le$39.76 & G & D & - & - & - & - & -\\
NGC1537 & 16.44 & 10.02 & 10.69 & 10.60 & $\le$39.75 & G & - & 0.291 & - & - & 2.06 & -\\
NGC1549 & 14.45 & 10.28 & 10.95 & 10.84 & 39.92 & G & - & 0.313 & -0.40 & - & - & -\\
NGC1550 & 48.49 & 10.33 & 11.21 & 11.10 & 42.80 & G & G & - & - & - & - & -\\
NGC1553 & 14.45 & 10.63 & 11.16 & 11.08 & 40.52 & G & G & 0.274 & - & 0.74 & - & -\\
NGC1573 & 51.52 & 10.72 & 11.35 & 11.23 & 41.33 & G & G & 0.331 & - & - & 1.85 & -\\
NGC1574 & 14.45 & 10.01 & 10.82 & 10.72 & $\le$40.32 & G & - & 0.287 & - & - & - & -\\
NGC1581 & 14.45 & 9.06 & 9.73 & 9.66 & $\le$39.86 & G & - & - & - & - & - & -\\
NGC1587 & 44.87 & 10.51 & 11.25 & 11.13 & 40.64 & G & - & 0.324 & - & - & 1.29 & -\\
NGC1600 & 59.98 & 11.03 & 11.69 & 11.57 & 41.54 & F & - & 0.329 & -0.75 & 0.08 & 1.87 & -6.62\\
NGC1705 & 4.87 & 8.44$^{*}$ & 8.51 & 8.42 & 38.81 & F & D & - & - & - & - & -\\
NGC1947 & 13.43 & 10.18 & 10.60 & 10.50 & $\le$40.05 & F & - & 0.243 & - & - & - & -\\
NGC2089 & 38.07 & 10.18$^{*}$ & 10.98 & 10.91 & $\le$40.56 & F & - & - & - & - & - & -\\
NGC2271 & 32.16 & 9.94$^{*}$ & 10.89 & 10.79 & $\le$40.66 & F & - & 0.306 & - & - & - & -\\
NGC2272 & 0.72 & 6.81 & 7.58 & 7.50 & $\le$37.36 & F & D & - & - & - & - & -\\
NGC2300 & 27.67 & 10.41 & 11.18 & 11.07 & 41.16 & G & - & 0.326 & 0.55 & - & 1.77 & -5.99\\
NGC2305 & 45.92 & 10.60 & 11.29 & 11.21 & 41.67 & F & - & 0.310 & - & - & - & -\\
NGC2314 & 48.53 & 10.44 & 11.17 & 11.06 & 40.91 & F & - & 0.319 & - & - & 1.69 & -\\
NGC2325 & 29.79 & 10.60 & 11.14 & 11.05 & 40.70 & F & - & 0.306 & - & - & 1.65 & -\\
NGC2328 & 12.20 & 9.02 & 9.55 & 9.47 & $\le$39.56 & F & - & - & - & - & - & -\\
NGC2329 & 71.12 & 10.73 & 11.29 & 11.20 & 42.12 & D & A & 0.269 & - & - & 0.97 & -\\
NGC2340 & 73.79 & 11.04 & 11.53 & 11.39 & 42.08 & D & C & 0.334 & - & - & 1.30 & -\\
NGC2380 & 21.11 & 9.93 & 11.02 & 10.89 & $\le$40.20 & F & - & 0.291 & - & - & - & -\\
NGC2434 & 14.06 & 9.89 & 10.49 & 10.39 & 39.90 & G & - & 0.275 & - & - & - & -\\
NGC2444 & 50.82 & 9.92 & 10.70 & 10.61 & $\le$41.29 & G & - & - & - & - & - & -\\
NGC2488 & 117.12 & 10.97 & 11.66 & 11.55 & 42.56 & D & C & - & - & - & - & -\\
NGC2502 & 11.07 & 9.00 & 10.29 & 10.18 & $\le$39.36 & F & - & 0.281 & - & - & - & -\\
NGC2563 & 59.43 & 10.54 & 11.29 & 11.19 & 41.63 & G & G & 0.328 & - & - & 1.08 & -\\
NGC2577 & 28.68 & 9.74 & 10.64 & 10.54 & 40.19 & F & - & - & - & - & - & -\\
NGC2629 & 52.13 & 10.29 & 11.24 & 11.13 & $\le$40.94 & G & - & 0.313 & - & - & 0.30 & -\\
NGC2634 & 33.49 & 10.07 & 10.69 & 10.61 & $\le$40.46 & G & - & 0.294 & - & 0.81 & 1.32 & -\\
NGC2663 & 27.42 & 10.95 & 11.50 & 11.38 & 40.45 & F & - & 0.335 & - & - & - & -\\
NGC2693 & 62.81 & 10.74 & 11.50 & 11.38 & $\le$41.29 & G & - & 0.332 & 1.00 & - & 1.15 & -\\
NGC2694 & 62.81 & 9.78 & 10.36 & 10.39 & $\le$41.31 & G & - & - & - & - & - & -\\
NGC2695 & 27.42 & 9.97 & 10.68 & 10.60 & $\le$40.39 & G & - & 0.319 & - & - & 1.05 & -\\
NGC2768 & 20.89 & 10.57 & 11.19 & 11.10 & 40.38 & G & G & 0.277 & - & - & 1.61 & -6.21\\
NGC2778 & 29.24 & 9.80 & 10.47 & 10.40 & $\le$40.33 & G & G & 0.317 & -0.15 & 0.94 & 1.24 & -\\
NGC2832 & 85.90 & 11.06 & 11.74 & 11.62 & 41.62 & D & G & 0.330 & -0.30 & 0.02 & 1.48 & -\\
NGC2865 & 36.48 & 10.48 & 11.09 & 10.99 & $\le$40.49 & F & - & 0.203 & - & - & 3.02 & -\\
NGC2880 & 23.55 & 9.95 & 10.62 & 10.50 & $\le$40.06 & G & - & - & - & - & - & -\\
NGC2887 & 35.01 & 10.17 & 11.23 & 11.11 & $\le$40.69 & F & - & 0.266 & - & - & - & -\\
NGC2888 & 27.12 & 9.62 & 10.44 & 10.36 & $\le$40.23 & F & - & 0.244 & - & - & - & -\\
NGC2904 & 29.35 & 9.75 & 10.60 & 10.49 & $\le$40.37 & F & - & 0.319 & - & - & - & -\\
NGC2911 & 41.30 & 10.52 & 11.10 & 11.00 & $\le$40.96 & G & G & - & - & - & 2.12 & -5.40\\
NGC2974 & 28.31 & 10.50 & 11.75 & 11.51 & 40.58 & G & G & 0.305 & 0.50 & - & 1.41 & -4.42\\
NGC2986 & 30.34 & 10.51 & 11.26 & 11.18 & 40.96 & F & - & 0.331 & - & 0.18 & 1.44 & -\\
NGC3065 & 30.03 & 9.74 & 10.71 & 10.62 & 41.01 & F & - & - & - & 0.79 & - & -\\
NGC3073 & 18.49 & 9.09 & 9.56 & 9.48 & $\le$39.77 & G & - & - & - & - & - & -\\
NGC3078 & 33.42 & 10.48 & 11.24 & 11.14 & 40.72 & G & - & 0.342 & - & 0.95 & 1.80 & -\\
NGC3087 & 34.67 & 10.52 & 11.17 & 11.06 & $\le$40.56 & G & - & 0.290 & - & - & 1.77 & -\\
NGC3091 & 50.82 & 10.75 & 11.52 & 11.43 & 41.63 & G & G & 0.327 & -0.50 & - & 1.43 & -6.77\\
NGC3115 & 8.83 & 10.10 & 10.89 & 10.80 & 39.72 & F & - & 0.326 & - & 0.78 & 1.85 & -6.97\\
NGC3136 & 19.11 & 10.07 & 11.06 & 10.94 & $\le$40.26 & F & - & 0.289 & - & - & - & -\\
NGC3156 & 19.95 & 9.71 & 10.13 & 10.03 & $\le$40.00 & G & - & 0.110 & - & - & 4.29 & 3.61\\
NGC3158 & 86.30 & 10.96 & 11.70 & 11.58 & 41.71 & D & C & 0.329 & - & - & 2.63 & -\\
NGC3193 & 21.58 & 10.15 & 10.83 & 10.75 & 39.96 & G & - & 0.299 & 0.30 & 0.81 & 1.66 & -6.18\\
NGC3224 & 38.55 & 10.16 & 10.98 & 10.89 & $\le$40.65 & G & - & - & - & - & - & -\\
NGC3226 & 21.58 & 10.12 & 10.59 & 10.47 & 40.20 & G & - & 0.299 & - & - & 1.01 & -6.07\\
NGC3250 & 37.67 & 10.71 & 11.42 & 11.32 & $\le$40.65 & G & - & 0.326 & -0.20 & - & - & -\\
NGC3258 & 38.37 & 10.48 & 11.19 & 11.10 & 41.17 & G & - & 0.354 & 0.00 & - & - & -\\
NGC3268 & 38.37 & 10.48 & 11.25 & 11.17 & 40.53 & G & - & 0.334 & - & - & - & -\\
NGC3271 & 47.81 & 10.43$^{*}$ & 11.45 & 11.34 & 41.06 & G & G & - & - & - & - & -\\
NGC3311 & 58.34 & 10.76 & 11.64 & 11.53 & 42.19 & C & C & 0.313 & - & - & - & -\\
NGC3375 & 30.92 & 9.80 & 10.42 & 10.35 & $\le$40.43 & F & - & 0.241 & - & - & 2.54 & -\\
NGC3377 & 10.00 & 9.72 & 10.37 & 10.30 & $\le$39.60 & G & - & 0.273 & 1.05 & 1.12 & 1.96 & -5.33\\
NGC3379 & 10.00 & 10.06 & 10.84 & 10.76 & $\le$39.54 & G & - & 0.318 & 0.10 & 0.18 & 1.50 & -6.63\\
NGC3384 & 10.00 & 9.85 & 10.65 & 10.55 & $\le$39.52 & G & - & 0.296 & - & 0.64 & 1.81 & -6.92\\
NGC3458 & 27.03 & 9.77$^{*}$ & 10.48 & 10.39 & $\le$40.61 & G & - & - & - & - & - & -\\
NGC3516 & 38.37 & 10.36 & 11.11 & 10.95 & 43.11 & G & A & - & - & - & - & -\\
NGC3557 & 32.21 & 10.76 & 10.89$^0$ & 10.89$^0$ & 40.58 & G & G & 0.317 & 0.00 & - & - & -\\
NGC3585 & 16.07 & 10.39 & 11.08 & 10.99 & 39.79 & G & G & 0.320 & - & - & 1.56 & -\\
NGC3599 & 19.77 & 9.66 & 10.23 & 10.15 & $\le$39.23 & G & G & 0.193 & - & 0.79 & 2.74 & -1.96\\
NGC3605 & 19.77 & 9.47 & 10.06 & 9.98 & 39.08 & G & - & 0.230 & -0.80 & 0.67 & 2.03 & -\\
NGC3606 & 37.75 & 9.98$^{*}$ & 10.75 & 10.66 & $\le$40.69 & F & - & 0.292 & - & - & - & -\\
NGC3607 & 19.77 & 10.46 & 11.14 & 11.03 & 40.54 & G & - & 0.313 & - & - & 1.92 & -6.13\\
NGC3608 & 19.77 & 10.11 & 10.70 & 10.59 & 40.01 & G & - & 0.319 & -0.20 & 0.00 & 1.61 & -6.46\\
NGC3610 & 27.29 & 10.40 & 11.06 & 10.96 & 39.83 & G & - & 0.269 & 2.50 & - & 2.33 & -4.85\\
NGC3613 & 27.29 & 10.36 & 11.00 & 10.93 & $\le$40.12 & G & G & 0.293 & 0.70 & 0.04 & 1.75 & -5.82\\
NGC3617 & 27.39 & 9.59 & 10.31 & 10.23 & $\le$40.40 & F & - & 0.205 & - & - & - & -\\
NGC3640 & 22.91 & 10.43 & 11.06 & 10.95 & 39.92 & G & G & 0.273 & -0.20 & 0.15 & 2.06 & -5.55\\
NGC3658 & 30.76 & 9.92 & 10.67 & 10.57 & $\le$39.77 & G & - & - & - & - & - & -\\
NGC3665 & 30.76 & 10.70 & 11.25 & 11.16 & 40.60 & G & G & 0.287 & - & - & 2.18 & -4.87\\
NGC3706 & 37.21 & 10.38 & 11.33 & 11.22 & $\le$41.19 & F & - & 0.315 & - & - & - & -\\
NGC3818 & 21.48 & 9.80 & 10.47 & 10.38 & $\le$40.16 & F & - & 0.330 & 2.30 & - & 1.69 & -\\
NGC3842 & 82.04 & 10.92 & 11.54 & 11.43 & 41.80 & D & C & 0.327 & -0.30 & - & 1.40 & -\\
NGC3862 & 82.04 & 10.56 & 11.38 & 11.27 & 41.90 & D & A & 0.291 & - & - & 1.15 & -\\
NGC3894 & 46.37 & 10.47 & 11.26 & 11.16 & 41.19 & G & - & 0.330 & -0.70 & - & 0.25 & -\\
NGC3904 & 17.86 & 10.06 & 10.78 & 10.68 & $\le$40.74 & G & - & 0.332 & - & - & 1.37 & -\\
NGC3923 & 17.86 & 10.52 & 11.25 & 11.15 & 40.66 & G & G & 0.324 & -0.40 & - & 1.92 & -6.86\\
NGC3962 & 21.68 & 10.28 & 10.95 & 10.87 & $\le$40.22 & F & - & 0.315 & 0.00 & - & 0.63 & -\\
NGC3990 & 12.16 & 8.99 & 9.70 & 9.62 & 38.60 & G & D & - & - & - & - & -\\
NGC3998 & 17.46 & 10.08 & 10.89 & 10.80 & 41.51 & G & A & 0.339 & - & - & - & -\\
NGC4024 & 20.84 & 9.77 & 10.49 & 10.43 & $\le$40.04 & G & - & 0.265 & - & - & 1.09 & -\\
NGC4033 & 19.25 & 9.71 & 10.44 & 10.35 & $\le$40.00 & G & - & 0.262 & - & - & 1.86 & -\\
NGC4036 & 21.73 & 10.24 & 11.00 & 10.91 & $\le$40.03 & G & G & 0.296 & - & - & - & -\\
NGC4073 & 79.43 & 11.07 & 11.75 & 11.67 & 42.38 & D & G & 0.325 & - & - & 1.72 & -\\
NGC4105 & 22.85 & 10.25$^{*}$ & 11.06 & 10.93 & 40.42 & G & G & 0.312 & - & - & 0.47 & -\\
NGC4125 & 25.94 & 10.80 & 11.43 & 11.34 & 40.94 & G & G & 0.311 & 0.95 & - & 1.76 & -5.82\\
NGC4168 & 33.73 & 10.40 & 11.03 & 10.93 & 40.56 & C & A & 0.266 & 0.37 & 0.14 & 1.26 & -\\
NGC4203 & 16.22 & 9.89 & 10.81 & 10.72 & 41.18 & G & A & 0.332 & - & - & - & -\\
NGC4233 & 31.48 & 10.02 & 10.83 & 10.73 & $\le$39.22 & C & - & 0.310 & - & - & - & -\\
NGC4239 & 16.75 & 9.24 & 9.72 & 9.65 & $\le$39.85 & C & - & 0.167 & - & 0.65 & 2.40 & -\\
NGC4251 & 16.22 & 10.02 & 10.68 & 10.57 & $\le$39.65 & G & - & 0.245 & 2.80 & - & - & -\\
NGC4261 & 31.48 & 10.70 & 11.44 & 11.34 & 41.21 & C & - & 0.344 & -1.30 & 0.00 & 1.41 & -6.57\\
NGC4262 & 15.92 & 9.65 & 10.41 & 10.32 & $\le$39.82 & C & - & 0.325 & - & - & 1.35 & -\\
NGC4267 & 15.92 & 9.88 & 10.62 & 10.53 & $\le$39.90 & C & - & - & - & - & - & -\\
NGC4278 & 16.22 & 10.24 & 10.89 & 10.81 & 40.36 & G & - & 0.306 & -1.00 & 0.02 & - & -\\
NGC4283 & 16.22 & 9.46 & 10.12 & 10.04 & $\le$39.22 & G & - & 0.281 & - & - & 1.78 & -\\
NGC4291 & 24.55 & 10.00 & 10.76 & 10.67 & 40.89 & G & - & 0.318 & -0.40 & 0.02 & 1.30 & -\\
NGC4339 & 15.92 & 9.71 & 10.34 & 10.26 & $\le$39.86 & C & - & 0.267 & - & - & 1.74 & -\\
NGC4350 & 15.92 & 9.85 & 10.63 & 10.55 & $\le$39.77 & C & - & 0.349 & - & - & 1.66 & -\\
NGC4365 & 15.92 & 10.34 & 11.10 & 11.01 & 40.25 & C & - & 0.328 & -0.95 & 0.15 & 1.73 & -6.55\\
NGC4374 & 15.92 & 10.57 & 11.26 & 11.17 & 40.83 & C & - & 0.316 & -0.40 & 0.31 & 1.84 & -6.05\\
NGC4386 & 24.55 & 9.90 & 10.73 & 10.65 & $\le$39.93 & G & - & - & - & - & - & -\\
NGC4387 & 15.92 & 9.47 & 10.09 & 10.02 & 39.71 & C & - & 0.252 & -0.75 & 0.72 & 1.61 & -\\
NGC4406 & 15.92 & 10.66 & 11.31 & 11.23 & 42.05 & C & C & 0.316 & -0.70 & 0.08 & 1.38 & -\\
NGC4417 & 15.92 & 9.77 & 10.48 & 10.39 & $\le$40.68 & C & - & 0.257 & - & 0.71 & - & -\\
NGC4434 & 15.92 & 9.45 & 10.07 & 9.98 & $\le$39.83 & C & - & 0.265 & 0.44 & 0.70 & 1.74 & -\\
NGC4435 & 15.92 & 10.01 & 10.83 & 10.89 & $\le$40.13 & C & - & 0.241 & - & - & - & -\\
NGC4458 & 16.14 & 9.51 & 10.04 & 9.94 & 39.84 & C & - & 0.233 & 0.34 & 0.49 & 1.91 & -\\
NGC4467 & 15.92 & 8.59 & 9.56 & 9.23 & $\le$39.29 & C & D & 0.270 & - & 0.98 & 1.59 & -\\
NGC4472 & 15.92 & 10.90 & 11.59 & 11.50 & 41.43 & C & C & 0.331 & -0.25 & 0.04 & 1.35 & -\\
NGC4473 & 16.14 & 10.15 & 10.90 & 10.81 & 40.14 & C & - & 0.310 & 0.90 & - & 1.30 & -\\
NGC4474 & 15.92 & 9.64 & 10.27 & 10.17 & $\le$39.85 & C & - & 0.245 & - & 0.72 & - & -\\
NGC4476 & 15.92 & 9.43 & 9.97 & 9.87 & $\le$40.27 & C & - & 0.164 & - & - & 2.35 & -\\
NGC4477 & 15.92 & 10.13 & 10.81 & 10.72 & 40.26 & C & - & - & - & - & - & -\\
NGC4478 & 15.92 & 9.79 & 10.41 & 10.33 & $\le$40.41 & C & - & 0.262 & -0.75 & 0.43 & 1.75 & -\\
NGC4479 & 15.92 & 9.23 & 9.84 & 9.75 & $\le$39.70 & C & - & 0.191 & - & - & - & -\\
NGC4486 & 15.92 & 10.85 & 11.43 & 11.33 & 42.95 & C & C & 0.304 & 0.00 & 0.25 & - & -\\
NGC4489 & 15.92 & 9.46 & 10.01 & 9.92 & $\le$39.84 & C & - & 0.207 & -0.20 & - & 2.16 & -\\
NGC4494 & 21.28 & 10.62 & 11.20 & 11.11 & $\le$40.10 & G & - & 0.287 & 0.30 & - & 1.77 & -5.99\\
NGC4503 & 15.92 & 9.77 & 10.60 & 10.50 & $\le$39.88 & C & - & - & - & 0.64 & - & -\\
NGC4515 & 15.92 & 9.24 & 9.80 & 9.73 & $\le$39.80 & C & - & 0.204 & - & - & - & -\\
NGC4526 & 15.92 & 10.47 & 11.16 & 11.08 & 39.87 & G & - & 0.304 & - & - & 1.42 & -\\
NGC4550 & 15.92 & 9.72 & 10.28 & 10.21 & 39.78 & C & - & 0.182 & 1.67 & - & 2.24 & -3.78\\
NGC4551 & 15.92 & 9.58 & 10.20 & 10.11 & $\le$39.09 & C & - & 0.266 & -0.65 & 0.80 & 1.59 & -\\
NGC4552 & 15.92 & 10.29 & 11.06 & 10.98 & 40.71 & C & A & 0.337 & 0.01 & 0.00 & 1.08 & -\\
NGC4555 & 90.33 & 10.86$^{*}$ & 11.59 & 11.49 & 41.85 & D & - & - & - & - & - & -\\
NGC4564 & 15.92 & 9.86 & 10.58 & 10.52 & $\le$39.85 & C & - & 0.332 & 1.00 & 0.05 & 1.65 & -\\
NGC4578 & 15.92 & 9.78 & 10.39 & 10.31 & $\le$39.99 & C & - & 0.310 & - & - & 1.02 & -\\
NGC4581 & 15.92 & 9.26 & 9.91 & 9.81 & $\le$39.96 & G & - & - & - & - & - & -\\
NGC4589 & 24.55 & 10.33 & 11.02 & 10.94 & 40.36 & G & G & 0.330 & 0.55 & 0.26 & 1.30 & -\\
NGC4621 & 15.92 & 10.32 & 11.05 & 10.97 & 40.02 & C & - & 0.336 & 1.50 & 0.50 & 1.60 & -\\
NGC4636 & 15.92 & 10.51 & 11.18 & 11.12 & 41.59 & G & G & 0.320 & -0.10 & 0.13 & 1.13 & -\\
NGC4638 & 15.92 & 9.80 & 10.47 & 10.37 & 39.59 & C & - & 0.272 & - & - & - & -\\
NGC4645 & 32.03 & 10.09$^{*}$ & 10.98 & 10.88 & $\le$40.57 & G & G & 0.287 & - & - & - & -\\
NGC4648 & 24.55 & 9.87 & 10.59 & 10.49 & $\le$39.89 & G & - & 0.329 & 0.00 & 0.92 & 1.38 & -\\
NGC4649 & 15.92 & 10.73 & 11.46 & 11.36 & 41.28 & C & - & 0.354 & -0.35 & 0.15 & 1.46 & -\\
NGC4660 & 15.92 & 9.74 & 10.47 & 10.38 & $\le$39.39 & C & - & 0.306 & 2.70 & - & 1.07 & -\\
NGC4697 & 15.14 & 10.55 & 11.16 & 11.07 & 40.12 & G & G & 0.308 & 1.30 & 0.74 & 1.08 & -\\
NGC4696 & 37.01 & 10.99$^{*}$ & 11.63 & 11.54 & 43.23 & C & C & 0.292 & - & - & - & -\\
NGC4709 & 59.31 & 10.94 & 11.69 & 11.58 & 41.00 & G & G & 0.336 & -0.80 & - & - & -\\
NGC4733 & 15.92 & 9.51 & 10.14 & 10.06 & $\le$39.73 & C & - & 0.196 & - & - & 2.46 & -\\
NGC4742 & 12.42 & 9.56 & 10.18 & 10.11 & $\le$39.80 & G & - & 0.193 & 0.41 & 1.09 & 3.32 & -\\
NGC4751 & 23.97 & 9.76$^{*}$ & 10.81 & 10.70 & $\le$40.31 & G & G & - & - & - & - & -\\
NGC4753 & 20.23 & 10.46 & 11.27 & 11.18 & 39.99 & G & G & - & - & - & - & -\\
NGC4754 & 15.92 & 10.00 & 10.79 & 10.70 & $\le$39.73 & C & - & 0.344 & - & - & 1.59 & -6.92\\
NGC4756 & 53.93 & 10.30 & 11.18 & 11.09 & 41.72 & G & - & - & - & - & - & -\\
NGC4760 & 63.39 & 11.02 & 11.53 & 11.43 & 41.58 & G & G & 0.305 & - & - & 1.28 & -\\
NGC4762 & 15.92 & 10.16 & 10.83 & 10.77 & 40.13 & C & - & 0.282 & - & - & 1.70 & -\\
NGC4767 & 37.39 & 10.33 & 11.20 & 11.10 & $\le$40.71 & C & - & 0.295 & - & - & - & -\\
NGC4782 & 63.39 & 11.37 & 11.85 & 11.74 & 41.61 & F & - & 0.341 & - & - & 2.05 & -\\
NGC4839 & 87.90 & 10.90 & 11.56 & 11.51 & 40.45 & D & - & 0.307 & - & - & 1.25 & -\\
NGC4889 & 88.31 & 11.19 & 11.88 & 11.76 & 42.76 & D & C & - & 0.01 & 0.05 & 1.14 & -\\
NGC4915 & 43.85 & 10.28 & 11.12 & 11.03 & $\le$40.87 & F & - & 0.294 & - & - & 1.55 & -\\
NGC4936 & 41.07 & 10.71$^{*}$ & 11.48 & 11.39 & 41.69 & G & G & 0.305 & - & - & - & -\\
NGC4946 & 38.53 & 9.98 & 10.96 & 10.86 & $\le$40.79 & G & - & 0.310 & - & - & - & -\\
NGC4976 & 11.43 & 9.97 & 10.73 & 10.62 & $\le$39.73 & G & - & 0.277 & 0.50 & - & - & -\\
NGC4993 & 37.49 & 10.01$^{*}$ & 10.81 & 10.70 & $\le$40.71 & G & - & - & - & - & - & -\\
NGC5011 & 38.76 & 10.40 & 11.27 & 11.17 & $\le$40.82 & G & - & 0.290 & - & - & - & -\\
NGC5018 & 30.20 & 10.57 & 11.21 & 11.10 & $\le$40.53 & G & G & 0.214 & - & - & 2.30 & -\\
NGC5044 & 30.20 & 10.70 & 11.22 & 11.12 & 42.74 & G & G & 0.329 & - & - & 0.06 & -\\
NGC5061 & 18.28 & 10.28 & 10.85 & 10.73 & 39.68 & G & G & 0.259 & - & - & 2.65 & -\\
NGC5077 & 30.20 & 10.26 & 11.02 & 10.91 & 40.48 & G & G & 0.307 & -0.90 & 0.23 & 0.84 & -\\
NGC5084 & 16.90 & 10.18 & 10.98 & 10.91 & 40.49 & G & G & 0.314 & - & - & 0.64 & -\\
NGC5087 & 18.71 & 10.03 & 10.77 & 10.68 & 40.36 & G & - & 0.328 & - & - & - & -\\
NGC5090 & 42.23 & 10.41 & 11.57 & 11.49 & 41.49 & G & A & 0.319 & - & - & - & -\\
NGC5102 & 4.16 & 9.29 & 9.82 & 9.78 & 38.03 & F & - & 0.0502 & - & - & 4.73 & -\\
NGC5128 & 3.89 & 10.45 & 10.95 & 10.82 & 40.10 & G & G & - & - & - & - & -\\
NGC5129 & 91.20 & 10.91 & 11.57 & 11.49 & 42.14 & D & G & 0.296 & - & - & 1.37 & -\\
NGC5153 & 55.65 & 10.55 & 11.18 & 11.09 & 40.50 & G & - & 0.296 & - & - & - & -\\
NGC5173 & 34.99 & 10.04 & 10.42 & 10.34 & $\le$40.36 & G & - & - & - & - & - & -\\
NGC5193 & 47.41 & 10.55 & 11.23 & 11.14 & 40.58 & G & - & 0.302 & - & - & - & -\\
NGC5198 & 34.99 & 10.28 & 10.88 & 10.79 & $\le$40.38 & G & G & 0.322 & - & 0.88 & 1.64 & -\\
NGC5216 & 42.33 & 10.02$^{*}$ & 10.65 & 10.55 & 40.85 & G & - & 0.265 & - & - & 0.73 & -\\
NGC5273 & 17.09 & 9.68 & 10.35 & 10.28 & 39.86 & F & A & - & - & 0.37 & - & -\\
NGC5306 & 96.41 & 10.91 & 11.73 & 11.62 & 41.50 & D & C & - & - & - & - & -\\
NGC5308 & 27.80 & 10.20 & 10.89 & 10.81 & $\le$40.01 & G & G & - & - & 0.82 & - & -\\
NGC5322 & 27.80 & 10.67 & 11.37 & 11.30 & 40.21 & G & - & 0.292 & -0.90 & - & 2.08 & -6.15\\
NGC5328 & 61.40 & 10.70 & 11.53 & 11.42 & 41.88 & G & - & 0.310 & - & - & 1.73 & -\\
NGC5353 & 34.67 & 10.56 & 11.38 & 11.34 & 41.00 & G & - & 0.324 & - & - & 1.68 & -7.08\\
NGC5382 & 58.19 & 10.32 & 11.00 & 10.94 & 40.14 & G & - & - & - & - & - & -\\
NGC5419 & 53.44 & 10.88 & 11.80 & 11.68 & 41.80 & G & G & 0.339 & - & - & - & -\\
NGC5473 & 28.18 & 10.21 & 10.90 & 10.80 & $\le$40.09 & G & - & - & - & - & - & -\\
NGC5485 & 28.18 & 10.25 & 10.89 & 10.80 & $\le$40.12 & G & G & 0.300 & - & - & 1.94 & -\\
NGC5507 & 25.85 & 9.63 & 10.53 & 10.44 & 39.75 & G & - & - & - & - & - & -\\
NGC5546 & 98.53 & 10.86 & 11.60 & 11.51 & 42.02 & D & C & - & - & - & - & -\\
NGC5574 & 21.68 & 9.57 & 10.20 & 10.11 & $\le$40.14 & G & - & - & - & - & - & -\\
NGC5576 & 21.68 & 10.16 & 10.89 & 10.84 & $\le$40.14 & G & - & 0.262 & -0.50 & 0.36 & 2.00 & -5.57\\
NGC5582 & 18.40 & 9.73 & 10.30 & 10.21 & $\le$39.82 & F & - & 0.294 & - & - & 1.41 & -\\
NGC5638 & 21.68 & 10.09 & 10.72 & 10.62 & $\le$40.20 & G & - & 0.322 & 0.20 & - & 1.61 & -6.86\\
NGC5687 & 31.49 & 10.15 & 10.76 & 10.69 & $\le$40.14 & F & - & 0.293 & - & - & 1.86 & -\\
NGC5812 & 24.55 & 10.19 & 10.94 & 10.85 & $\le$40.32 & F & - & 0.326 & 0.00 & 0.59 & 1.63 & -6.97\\
NGC5831 & 22.91 & 10.03 & 10.69 & 10.60 & $\le$40.25 & G & - & 0.297 & 0.50 & 0.33 & 1.47 & -5.63\\
NGC5838 & 22.91 & 10.20 & 11.04 & 10.96 & 40.02 & G & - & - & - & 0.93 & - & -\\
NGC5845 & 22.91 & 9.56 & 10.42 & 10.33 & $\le$39.94 & G & - & 0.315 & 0.72 & 0.51 & 1.93 & -6.05\\
NGC5846 & 22.91 & 10.66 & 11.29 & 11.20 & 41.65 & G & G & 0.328 & - & - & 1.52 & -7.03\\
NGC5898 & 23.88 & 10.22 & 10.88 & 10.78 & $\le$40.31 & G & - & 0.313 & - & 0.41 & - & -\\
NGC5903 & 23.88 & 10.28 & 10.86 & 10.77 & $\le$40.33 & G & G & 0.288 & -0.70 & 0.00 & - & -\\
NGC5982 & 37.50 & 10.53 & 11.24 & 11.16 & 41.16 & G & - & 0.293 & -0.80 & 0.11 & 1.47 & -5.86\\
NGC6034 & 137.28 & 10.63 & 11.44 & 11.33 & 42.19 & D & A & - & - & - & - & -\\
NGC6127 & 65.01 & 10.61 & 11.32 & 11.22 & 41.40 & F & - & 0.314 & - & - & 1.31 & -\\
NGC6137 & 112.72 & 11.13 & 11.63 & 11.53 & 42.14 & D & - & 0.289 & - & - & 0.61 & -\\
NGC6146 & 107.65 & 10.92 & 11.62 & 11.52 & $\le$41.90 & D & - & 0.280 & - & - & - & -\\
NGC6160 & 127.65 & 10.72$^{*}$ & 11.65 & 11.56 & 42.45 & D & - & 0.302 & - & - & 1.66 & -\\
NGC6173 & 107.65 & 11.09 & 11.67 & 11.57 & 42.15 & D & C & 0.314 & - & - & - & -\\
NGC6269 & 139.68 & 11.15 & 11.88 & 11.76 & 42.86 & D & C & 0.339 & - & - & - & -\\
NGC6305 & 33.72 & 9.95 & 10.85 & 10.75 & $\le$40.94 & F & - & 0.251 & - & - & - & -\\
NGC6407 & 58.67 & 10.58$^{*}$ & 11.42 & 11.35 & 41.94 & G & G & 0.313 & - & - & - & -\\
NGC6482 & 54.69 & 10.72 & 11.48 & 11.38 & 42.08 & F & - & 0.328 & 2.50 & - & 0.44 & -\\
NGC6487 & 105.39 & 11.07 & 11.69 & 11.58 & 41.70 & D & C & - & - & - & - & -\\
NGC6673 & 12.29 & 9.34 & 10.24 & 10.15 & $\le$40.06 & F & - & 0.257 & - & - & - & -\\
NGC6684 & 8.47 & 9.73 & 10.38 & 10.35 & 38.93 & F & - & 0.246 & - & - & - & -\\
NGC6703 & 29.92 & 10.37 & 11.00 & 10.90 & $\le$40.03 & F & - & 0.285 & 0.00 & - & 1.72 & -\\
NGC6776 & 70.41 & 10.66 & 11.43 & 11.29 & 40.79 & D & - & 0.243 & - & - & - & -\\
NGC6841 & 0.15 & 5.01 & 6.14 & 6.03 & $\le$36.06 & F & D & - & - & - & - & -\\
NGC6851 & 34.67 & 10.30 & 10.92 & 10.83 & $\le$40.64 & G & - & 0.270 & - & - & - & -\\
NGC6861 & 34.67 & 10.42 & 11.34 & 11.24 & $\le$40.65 & G & - & - & - & - & - & -\\
NGC6868 & 34.67 & 10.58$^{*}$ & 11.50 & 11.41 & 41.23 & G & G & 0.331 & 0.50 & - & - & -\\
NGC6876 & 48.56 & 10.83$^{*}$ & 11.64 & 11.54 & 41.51 & G & G & 0.290 & 0.60 & - & - & -\\
NGC6909 & 34.67 & 10.27 & 10.74 & 10.68 & $\le$40.78 & F & - & 0.213 & -0.90 & - & - & -\\
NGC6920 & 34.13 & 9.83 & 11.07 & 10.95 & $\le$40.95 & F & - & - & - & - & - & -\\
NGC6958 & 34.79 & 10.34 & 11.07 & 10.97 & $\le$40.68 & F & - & 0.239 & - & - & - & -\\
NGC6964 & 51.83 & 10.02 & 10.96 & 10.84 & $\le$40.84 & F & - & - & - & - & - & -\\
NGC7007 & 37.39 & 10.16 & 10.91 & 10.82 & $\le$40.70 & F & - & - & - & - & - & -\\
NGC7029 & 34.97 & 10.34 & 11.02 & 10.95 & $\le$40.57 & F & - & 0.255 & 2.80 & - & - & -\\
NGC7041 & 23.39 & 10.09 & 10.80 & 10.70 & $\le$40.24 & G & - & - & - & - & - & -\\
NGC7049 & 27.27 & 10.37 & 11.32 & 11.19 & 41.01 & G & G & - & - & - & - & -\\
NGC7097 & 29.24 & 10.13 & 10.81 & 10.71 & 40.28 & G & - & 0.305 & - & - & - & -\\
NGC7144 & 21.38 & 10.14 & 10.82 & 10.77 & 39.64 & G & G & 0.285 & - & - & - & -\\
NGC7145 & 21.38 & 10.04 & 10.61 & 10.53 & $\le$40.25 & G & - & 0.259 & - & - & - & -\\
NGC7166 & 30.43 & 10.02 & 10.91 & 10.83 & $\le$40.45 & G & - & - & - & - & - & -\\
NGC7168 & 34.57 & 10.12 & 10.87 & 10.77 & $\le$40.59 & F & - & - & - & - & - & -\\
NGC7173 & 32.65 & 10.04 & 10.79 & 10.71 & 40.86 & G & - & 0.289 & - & - & 0.88 & -\\
NGC7176 & 32.39 & 10.27 & 11.16 & 11.06 & 40.80 & G & - & 0.323 & - & - & 1.20 & -\\
NGC7180 & 16.05 & 9.18 & 9.91 & 9.81 & $\le$40.07 & F & - & 0.209 & - & - & 0.95 & -\\
NGC7185 & 24.04 & 9.59 & 10.23 & 10.20 & $\le$40.38 & F & - & - & - & - & - & -\\
NGC7192 & 35.77 & 10.42 & 11.05 & 10.93 & 40.85 & G & G & 0.261 & - & - & - & -\\
NGC7196 & 36.51 & 10.35 & 11.16 & 11.04 & 40.95 & F & - & 0.293 & - & - & - & -\\
NGC7236 & 105.60 & 10.39 & 11.45 & 11.30 & $\le$41.63 & D & - & 0.295 & - & - & 1.58 & -\\
NGC7252 & 52.48 & 10.66 & 11.06 & 10.96 & 40.50 & G & - & - & - & - & - & -\\
NGC7265 & 68.17 & 10.56 & 11.54 & 11.44 & 41.70 & F & - & - & - & - & - & -\\
NGC7332 & 15.28 & 9.86 & 10.51 & 10.42 & $\le$40.01 & F & - & 0.242 & - & 0.90 & 2.49 & -5.09\\
NGC7454 & 24.32 & 9.95 & 10.58 & 10.48 & $\le$40.20 & G & - & 0.215 & - & - & 2.21 & -4.66\\
NGC7457 & 10.67 & 9.78 & 10.13 & 10.02 & $\le$39.49 & F & - & 0.178 & 0.00 & 0.35 & 2.45 & -\\
NGC7465 & 24.32 & 9.64 & 10.30 & 10.19 & 41.36 & G & A & - & - & - & - & -\\
NGC7484 & 34.69 & 10.16$^{*}$ & 10.89 & 10.78 & $\le$40.93 & F & - & - & - & - & - & -\\
NGC7507 & 17.78 & 10.23 & 10.93 & 10.85 & $\le$40.77 & F & - & 0.345 & - & - & 1.42 & -\\
NGC7550 & 69.64 & 10.61 & 11.47 & 11.36 & $\le$40.31 & G & G & - & - & - & - & -\\
NGC7562 & 39.99 & 10.46 & 11.22 & 11.12 & $\le$40.93 & C & - & 0.292 & - & - & 1.72 & -\\
NGC7619 & 39.99 & 10.58 & 11.34 & 11.21 & 41.63 & C & C & 0.345 & 0.30 & - & 1.48 & -6.89\\
NGC7626 & 39.99 & 10.61 & 11.34 & 11.25 & 41.06 & C & C & 0.339 & - & 0.36 & 1.33 & -6.47\\
NGC7768 & 92.04 & 10.92 & 11.54 & 11.43 & 41.74 & D & C & 0.312 & 0.00 & 0.00 & 0.83 & -\\
NGC7796 & 39.45 & 10.48 & 11.23 & 11.12 & 40.74 & F & - & 0.254 & - & - & - & -\\
UGC1308 & 55.21 & 10.16$^{*}$ & 11.13 & 11.07 & 40.98 & G & - & 0.315 & - & - & - & -\\
UGC4956 & 67.63 & 10.45 & 11.14 & 11.05 & 41.60 & C & - & 0.321 & - & - & - & -\\

\end{longtable}
\twocolumn

\end{document}